%% file: main.tex
\documentclass[
        authorcolumns,
        USenglish,
    ]{lipics-v2021}

\usepackage[
    ]{auxlib}
\usepackage{float}

\usepackage[
        numbers,
        sort&compress,
    ]{natbib}
\makeatletter
\newcommand\savetreesparagraph{\@startsection{subparagraph}{5}{\z@}%
                                       {1.625ex \@plus0.5ex \@minus .1ex}%
                                       {-1em}%
                                      {\sffamily\normalsize\bfseries}}
\def\paragraph#1{\savetreesparagraph{#1.}}
   \abovedisplayshortskip \z@ \@plus3\p@
   \belowdisplayskip \abovedisplayskip
\makeatother

\usepackage{wasysym}

\RenewDocumentEnvironment{alphaenumerate}{O{}}{\enumerate[(a),#1]}{\endenumerate}
\RenewDocumentEnvironment{bracketenumerate}{O{}}{\enumerate[(1),#1]}{\endenumerate}
\RenewDocumentEnvironment{romanenumerate}{O{}}{\enumerate[(i),#1]}{\endenumerate}

\let\dcmciteauthor\citeauthor
\def\citeauthor#1{{\hypersetup{citecolor=black}\dcmciteauthor{#1}}}

\createcustomnote{todo}{backgroundcolor=AUXLorange, prefix=TODO}

\createcustomnote{christoph}{backgroundcolor=AUXLyellow}
\createcustomnote{dominik}{backgroundcolor=AUXLlightred}
\createcustomnote{florian}{backgroundcolor=AUXLlightblue}
\createcustomnote{peter}{backgroundcolor=AUXLlightgreen}

\renewcommand*{\todo}[2]{}
\renewcommand*{\christoph}[2]{}
\renewcommand*{\dominik}[2]{}
\renewcommand*{\florian}[2]{}
\renewcommand*{\peter}[2]{}

\def\LLARPlong/{\textsc{Lower-Left Anchored Rectangle Packing}}
\def\LLARP/{LLARP}

\def\GreedyPacking/{\textsc{GreedyPacking}}
\def\TilePacking/{\textsc{TilePacking}}

\newcommand*{\LBi}{\xi_w}
\newcommand*{\LBii}{\xi_s}

\DeclareMathOperator{\arcsinh}{arsinh}
\DeclareRobustCommand{\ammaG}{\text{%
    $\Gamma$%
}}

\makeatletter
\newcommand*{\rom}[1]{\expandafter\@slowromancap\romannumeral #1@}
\makeatother

\newcommand*{\TypedLine}[2]{\ell^{#1}_{#2}}

\newcommand*{\aIndicator}{\phantomas[c]{-}{\rotatebox[origin=c]{ 45}{$-$}}}
\newcommand*{\aLine}[1]{\TypedLine{\aIndicator}{#1}}

\newcommand*{\dIndicator}{\phantomas[c]{-}{\rotatebox[origin=c]{-45}{$-$}}}
\newcommand*{\dLine}[1]{\TypedLine{\dIndicator}{#1}}

\newcommand*{\hIndicator}{-}
\newcommand*{\hLine}[1]{\TypedLine{-}{#1}}

\newcommand*{\vIndicator}{\phantomas[c]{-}{\rotatebox[origin=c]{ 90 }{$-$}}}
\newcommand*{\vLine}[1]{\TypedLine{\vIndicator}{#1}}

\newcommand*{\hyperbola}[1]{h_{#1}}

\newcommand*{\wcTileLD}[1]{t_{h}(#1)}
\newcommand*{\wcTileHD}[1]{t_{l}(#1)}

\allowdisplaybreaks

\def\OurFancyTitle{
    On Greedily Packing Anchored Rectangles
}
\def\OurFancySubtitle{
    \Large A tale of crowns and tiles
}
\title{\OurFancyTitle\texorpdfstring{\\\OurFancySubtitle}{}}

\titlerunning{\OurFancyTitle}

\author{Christoph Damerius}{Universität Hamburg, Germany}{christoph.damerius@uni-hamburg.de}{}{}
\author{Dominik Kaaser}{Universität Hamburg, Germany}{dominik.kaaser@uni-hamburg.de}{https://orcid.org/0000-0002-2083-7145}{}
\author{Peter Kling}{Universität Hamburg, Germany}{peter.kling@uni-hamburg.de}{https://orcid.org/0000-0003-0000-8689}{}
\author{Florian Schneider}{Universität Hamburg, Germany}{fschneider@informatik.uni-hamburg.de}{}{}

\authorrunning{C.\ Damerius, D.\ Kaaser, P.\ Kling, and F.\ Schneider}


\nolinenumbers

\hideLIPIcs

\modulolinenumbers[5]

\begin{document}
\maketitle

{\input{00-abstract}}

{\input{10-introduction}}
{\input{20-problem-algorithm-description}}
{\input{30-general-approach}}
{\input{40-charging-scheme-and-weak-covering}}
{\input{50-strong-covering}}
{\input{60-upper-bound}}

\clearpage

\def\doi#1{doi: \href{https://doi.org/#1}{#1}}
\bibliography{references}

\begin{figure}[b!]
\centering
\includegraphics[width=\linewidth]{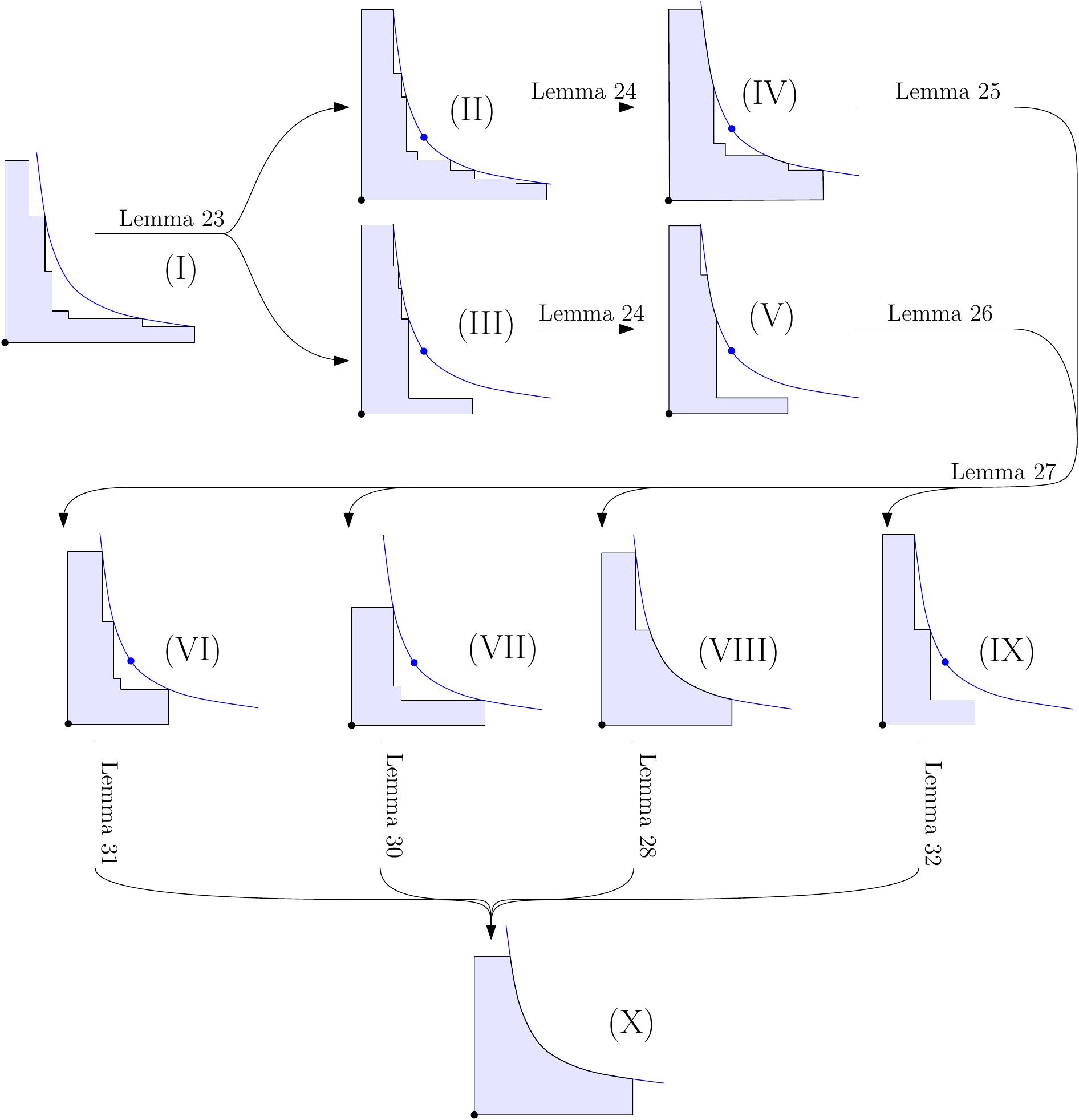}
\caption{
    Transforming low-density tiles $t$ with $\rho_t \leq 1/2$ to a corresponding worst-case hyperbola tile $\wcTileLD{s} \preceq t$.
    For the normalized tiles in Cases (\rom{2}) and later, the blue dot marks the point $(1, 1)$.
}\label{fig:transformation_plan}
\end{figure}

\clearpage
\appendix
\crefalias{section}{appendix}


{\input{90-appendix}}
{\input{91-upper-bound-full}}
\end{document}

%% file: 00-abstract.tex
\begin{abstract}
Consider a set $P$ of points in the unit square $\cU$, one of them being the origin.
For each point $p \in P$ you may draw a rectangle in $\cU$ with its lower-left corner in $p$.
What is the maximum area such rectangles can cover without overlapping each other?

Freedman~\cite{Freedman69} posed this problem in 1969, asking whether one can always cover at least $50\%$ of $\cU$.
Over 40 years later, \textcite{DBLP:journals/combinatorica/DumitrescuT15} achieved the first constant coverage of $9.1\%$; since then, no significant progress was made.
While $9.1\%$ might seem low, the authors could not find any instance where their algorithm covers less than $50\%$, nourishing the hope to eventually prove a $50\%$ bound.
While we indeed significantly raise the algorithm's coverage to $39\%$, we extinguish the hope of reaching $50\%$ by giving points for which the coverage is below $43.3\%$.

Our analysis studies the algorithm's average and worst-case density of so-called tiles, which represent the area where a given point can freely choose its maximum-area rectangle.
Our approach is comparatively general and may potentially help in analyzing related algorithms.
\end{abstract}

%% file: 10-introduction.tex
\section{Introduction}%
\label{sec:introduction}

The \LLARPlong/ (\LLARP/) problem considers a finite set $P \subseteq \cU \coloneqq \intco{0, 1}^2$ with $(0, 0) \in P$ of \emph{input points}.
The goal is to find a set of non-empty, interior-disjoint rectangles $\intoo{r_p}_{p \in P}$ with $p$ being the lower-left corner of $r_p \subseteq \cU$ and such that the total area $\sum_{p \in P} \abs{r_p}$ is maximized.

This problem was first introduced by Freedman~\cite[Unsolved Problem~11, page~345]{Freedman69} in 1969, who asked the question whether for any point set $P$, the rectangles can always be chosen such that they cover at least $50\%$ of $\cU$.
It is easy to see that this is the best one can hope for, since putting $n$ equally spaced points along the ascending diagonal of $\cU$ yields a maximum coverable area of $1/2 + \ldauomicron{1}$ for $n \to \infty$.

Over the years, the \LLARP/ problem reoccurred in form of geometric challenges~\cite{IBMPonderThis} or in miscellaneous books and journals about mathematical puzzles~\cite{Winkler, 10.1145/1839676.1839700, 10.1145/1810891.1810917}.
Still, it took more than 40 years until \textcite{DBLP:journals/combinatorica/DumitrescuT15} made significant progress towards Freedman's question: they considered a natural greedy algorithm and proved that it achieves a coverage of $9.1\%$.
This caused a surge of interest in this old problem, resulting in numerous findings for variants or special cases of the problem (see \cref{sec:related_work}).
Still, no further significant progress was made towards the original question\footnote{%
    A very recent, still unpublished result slightly raised the greedy algorithm's coverage to $10.39\%$~\cite{hoeksma2021better}.
}, and even the question whether a maximum covering can be found in polynomial time remains elusive.

While \textcite{DBLP:journals/combinatorica/DumitrescuT15} themselves observed that \enquote{\emph{a sizable gap to the conjectured $50\%$ remains}}, they were unable to find instances where it \emph{does not} reach $50\%$.
This led them and others to conjecture a much better quality of their algorithm, making it a natural candidate to answer Freedman's question positively, albeit \cite{DBLP:journals/combinatorica/DumitrescuT15} also mentioned that \enquote{\emph{obtaining substantial improvements probably requires new ideas}}.

Our results indeed attest the greedy algorithm a much better coverage of $39\%$.
However, at the same time we show that there are instances where the coverage stays below $43.3\%$.

\subsection{Related Work}%
\label{sec:related_work}

\LLARP/ falls into the class of geometric packing problems, where a typical question is how much of a container can be covered using a set of geometric shapes in two or more dimensions.
We concentrate on two dimensional packing problems with rectangular containers and shapes.

\paragraph{Complexity of \LLARP/}
The above mentioned greedy algorithm by \textcite{DBLP:journals/combinatorica/DumitrescuT15} considers the input points step by step from top-right to bottom-left, greedily choosing maximum-area rectangles in each step.
It can be shown~\cite{DBLP:journals/combinatorica/DumitrescuT15} that this is equivalent to partitioning the unit square into staircase-shaped \emph{tiles}, one for each input point, and choosing maximal rectangles within each tile (see \cref{sec:preliminaries} for the formal algorithm description).

While the complexity of \LLARP/ remains unknown, \cite{DBLP:journals/combinatorica/DumitrescuT15} also showed that there is an order of the input points for which the greedy algorithm achieves an optimal packing (albeit of unknown value); how to find that ordering remains unclear.
\Textcite{DBLP:journals/tcs/BalasT16} studied the combinatorial structure of optimal solutions, proving that the worst-case number of maximal rectangle packings is exponential in the number of input points.

\paragraph{\LLARP/ Variants}
After~\cite{DBLP:journals/combinatorica/DumitrescuT15}, a series of papers studied special cases and variants of \LLARP/.
\Textcite{DBLP:journals/disopt/BalasDT17} allowed rectangles to be anchored in any of the four corners and showed that here the worst-case coverage lies in $\intcc{7/12, 2/3}$ and in $\intcc{5/32, 7/23}$ if the rectangles are restricted to squares.
\Textcite{DBLP:conf/mfcs/AkitayaJST18} improved the lower bound for such corner-anchored square packings to $1/2$ and proved that finding such a maximum packing is NP-hard.\christoph{the $1/2$ corresponds to the \emph{reach}, i.e., the area if you take the union of all (possibly overlapping) squares, not the coverage.}
Interestingly, there is only one other \LLARP/-variant known to be NP-hard, namely if the rectangle's anchors lie in their center~\cite{DBLP:conf/esa/AntoniadisBCDHK19}.
Other results consider specific classes of input points, like points with certain increasing/decreasing structures~\cite{bian2018special} or points that lie on the unit square's boundary (for corner-anchored rectangles)~\cite{DBLP:journals/comgeo/BiedlBMM20}.

\paragraph{Further Problems and Applications}
Further related problems include the maximum weight independent set of rectangles problem~\cite{DBLP:conf/focs/AdamaszekW13, DBLP:journals/dcg/ChanH12, DBLP:conf/approx/AdamaszekCW15} (which was used, e.g., in~\cite{DBLP:conf/esa/AntoniadisBCDHK19} to derive a PTAS for center-anchored rectangle packings) or geometric knapsack~\cite{DBLP:conf/soda/AdamaszekW15, DBLP:conf/icalp/MerinoW20} and strip packing problems~\cite{DBLP:journals/tcs/JansenR19}.
All of these differ from \LLARP/ and its variants in that the size of the objects to be packed is part of the input and the object placement is typically less constrained.

Note that \LLARP/-like problems are not of pure theoretical interest, but have applications in, e.g., map labeling.
Here, rectangular text labels must be placed under certain constraints (e.g., labels might be scalable but require a fixed ratio and must be placed at a specific anchor) within a given container.
We refer to the relatively recent survey~\cite{DBLP:reference/crc/KakoulisT13} for details.

\subsection{Our Contribution and Techniques}%
\label{sec:contribution}

We analyze the greedy \TilePacking/ from~\cite{DBLP:journals/combinatorica/DumitrescuT15} (formally described in \cref{sec:preliminaries}).
From a high-level view, \TilePacking/ partitions the unit square into staircase-shaped \emph{tiles}, each anchored at an input point, and chooses an area-maximal rectangle in each tile.
A natural way to analyze such an algorithm is to consider tiles' densities (the ratio between their area-maximal rectangles and their own area) and prove a lower bound on the \emph{average} tile density (which immediately yields the covering guarantee).
\christoph{following sentence is weird. low-density tiles do not use up a comparatively large space...}
Indeed, intuitively there cannot be too many low-density tiles, as such tiles \enquote{use up} a comparatively large space.

\Textcite{DBLP:journals/combinatorica/DumitrescuT15} follow this approach by defining suitable charging areas $C_t$  for each tile $t$ (suitable trapezoids below/beneath the tile).
We also use such a charging scheme, but rely on a much more complex charging area which we refer to as a tile's \emph{crown}.
But instead of directly analyzing a tile's charging area, we first extract the critical properties that determine the charging scheme's quality.
This general approach (described in \cref{sec:general_approach}) requires a lower bound function $\xi$ on the tile's \emph{charging ratio} $\abs{C_t} / \abs{t}$ together with some simple properties (basically a form of local convexity characterizing the average tile density).

We derive such a lower bound and describe simple, symmetric tiles for which it is tight (\cref{fig:extreme_case_low_density, fig:extreme_case_high_density}).
We then take an arbitrary tile and show how to gradually transform it into one of these tiles without increasing its charging ratio.
This established\christoph{establishes?} that $\xi$ is indeed a lower bound and allows us to easily conclude the following \lcnamecref{thm:strong_covering_guarantee}.
\begin{theorem}[{name=, restate=[name=]thmStrongCoveringGuarantee}]%
\label{thm:strong_covering_guarantee}
For any \christoph{set of?} input points, \TilePacking/ covers at least $39\%$ of the unit square.
\end{theorem}

While the involved transformations to get from an arbitrary tile to a worst-case tile require some care, we showcase the versatility of our approach by first proving a slightly weaker bound of only $25\%$ (\cref{sec:weak_covering_guarantee}).
The analysis of this bound is not only much simpler but, in fact, takes us halfway to \cref{thm:strong_covering_guarantee}, as the $\xi$-bound used in this case (\cref{prop:LBi}) is tight for high-density tiles and all that remains is to refine our bound for low-density tiles (\cref{prop:LBii}).

Our second major result constructs an input instance (depicted in \cref{fig:upper_bound_simplified}) for which \TilePacking/ covers significantly less than $50\%$.
\begin{theorem}[{name=, restate=[name=restated]thmWorstCaseCovering}]%
\label{thm:worst_case_covering}
There is a set of input points $P$ for which algorithm \TilePacking/ covers at most $43.3\%$ of the unit square.
\end{theorem}

%% file: 20-problem-algorithm-description.tex
\section{Preliminaries and Algorithm Description}%
\label{sec:preliminaries}

Let $\cU \coloneqq \intco{0, 1}^2$ denote the unit square.
For a point $p \in \R^2$ define $x(p)$ and $y(p)$ as the $x$- and $y$-coordinates of $p$, respectively.
For two points $p, p' \in \R^2$ we use the notation $p \preceq p'$ to indicate that $x(p) \leq x(p')$ and $y(p) \leq y(p')$.
Similarly, $p \prec p'$ means that $x(p) < x(p')$ and $y(p) < y(p')$.
The relations \enquote{$\succeq$} and \enquote{$\succ$} are defined analogously.
For a set $S$ we denote its closure by $\overline{S}$.
If $S$ is measurable, we use $\abs{S}$ to denote its area.

To simplify some geometric constructions and arguments, we use the following line-notation:
We define the line $\aLine{q} \subseteq \R^2$ as the line through $q \in \R^2$ of slope $+1$.
Similarly, we define the lines $\hLine{q}$, $\dLine{q}$, and $\vLine{q}$ through $q$ with slope $0$, $-1$, and $\infty$, respectively.
For lines of type $R \in \set{\aIndicator, \vIndicator}$ we write $\TypedLine{R}{q} < \TypedLine{R}{q'}$ if $q' = q + (x, 0)$ with $x > 0$ and say $\TypedLine{R}{q}$ \emph{is left of} $\TypedLine{R}{q'}$.
Similarly, for lines of type $R \in \set{\dIndicator, \hIndicator}$ we write $\TypedLine{R}{q} < \TypedLine{R}{q'}$ if $q' = q + (0,y)$ with $y > 0$ and say $\TypedLine{R}{q}$ \emph{is below} $\TypedLine{R}{q'}$.
Analogous definitions apply for \enquote{$>$}, \enquote{$\leq$}, and \enquote{$\geq$}.

\paragraph{Input Sets in General Position}
Remember the problem description from \cref{sec:introduction}.
We say that the input set $P$ is \emph{in general position} if there are no two (different) points $p, p' \in P$ with $x(p) = x(p')$, $y(p) = y(p')$, or $x(p) + y(p) = x(p') + y(p')$.
That is, no two points may share an $x$- or $y$-coordinate and may not lie on the same diagonal of slope $-1$.
W.l.o.g., we restrict $P$ to be in general position (\cref{lem:general_point_sets} in \cref{app:auxiliary} explains why this is okay).

\begin{figure}
\vspace{-2ex}
\begin{subfigure}[b]{120pt}
\includegraphics{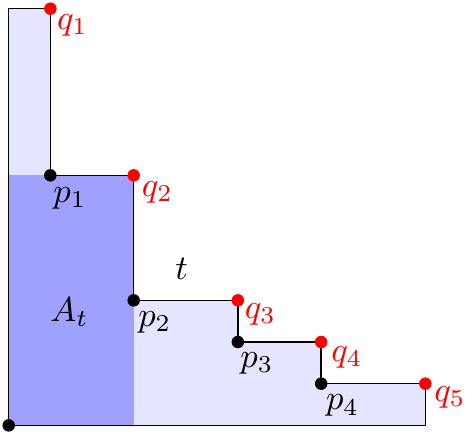}
\caption{Staircase points $p_i$, $q_j$ and rectangle $A_t$.}
\label{fig:tile_staircase_points}
\end{subfigure}
\hfill
\begin{subfigure}[b]{120pt}
\includegraphics{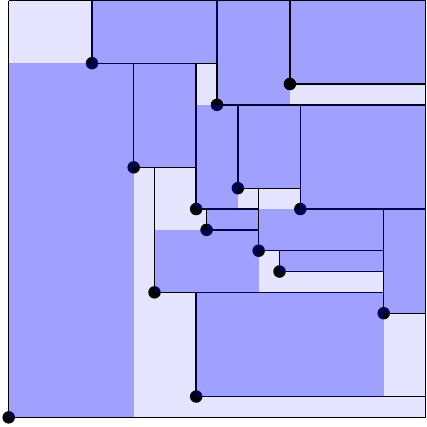}
\caption{A packing produced by \textsc{TilePacking}.}
\label{fig:tile_packing}
\end{subfigure}
\hfill
\begin{subfigure}[b]{120pt}
\makebox[120pt][c]{\includegraphics{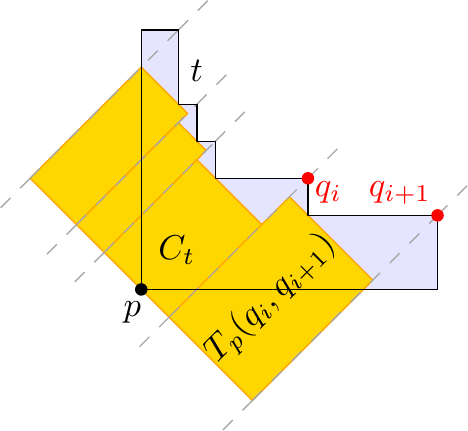}}
\caption{A tile $t$, its crown $C_t$, and a tower $T_p(q_i,q_{i+1})$.}
\label{fig:tile_crown}
\end{subfigure}
\caption{Tiles, packings, crowns, and towers.
In our figures, tiles are shaded light blue. Upper and lower stair case points are shown in red and black, respectively. A
dark blue rectangle represents a (maximal) rectangle of a tile.
Crowns are shown in yellow and towers are possibly labeled.
}
\end{figure}

\paragraph{Tiles and Tile Packings}
A \emph{tile} $t \subseteq \cU$ is a staircase polygon in $\cU$ (see \cref{fig:tile_staircase_points}).
More formally, $t$ is defined by its \emph{anchor} $p \in \cU$ and a set of $k$ \emph{upper staircase points} $\Gamma_t \coloneqq \set{q_1, q_2, \dots, q_k} \subseteq \overline{\cU}$ ordered by increasing $x$-coordinate and such that $q_i \succ p$ for all $q_i$ as well as $q_i \not\preceq q_j$ for all $q_i \neq q_j$.
Then $t = \set{q \in \cU | q \succeq p \land \exists q' \in \Gamma_t\colon q \prec q'}$.
A point $p_i = \bigl(x(q_{i-1}), y(q_i)\bigr)$ is called a \emph{lower staircase point}.
We define $A_t \subseteq t$ as an (arbitrary) area-maximal rectangle in $t$ and $\rho_t \coloneqq \abs{A_t} / \abs{t}$ as the tile's \emph{density}.
For indexed upper staircase points $q_i$ we often use the shorthands $x_i \coloneqq x(q_i)$ and $y_i \coloneqq y(q_i)$.

If $p$ and $\Gamma_t$ do not adhere to the above requirements ($q_i \succ p$ and $q_i \not\preceq q_j$), but only to the slightly weaker requirements $q_i \succeq p$ $q_i \not\preceq q_j$ for all $q_i \neq q_j$), then tile $t$ is said to be \emph{degenerate}.
Such tiles have superfluous points in $\Gamma_t$ and play a role in our analysis.

\christoph{hyperbola definition is wrong. We are adding a point to an $x$-value? $|A_t|/x$ should also be something like $|A_t|/(x-x(p))$ or something like that. Alternatively, use $\set{(x+x(p), y+y(p)) \in \R_{>0}^2 | y = \abs{A_t}/x}$}{}
The \emph{hyperbola} of $t$ is $\hyperbola{t} \coloneqq \set{(x, y) \in \R_{>0}^2 | y = p + \abs{A_t}/x}$.
Note that all upper staircase points lie between $p$ and $\hyperbola{t}$.
Moreover, the points from $\Gamma_t \cap h_t$ span all area-maximal rectangles in $t$.
If, $p = (0, 0)$ and $\abs{A_t} = 1$, then $t$ is called \emph{normalized}.

A \emph{tile packing} of the unit square is a set $\cT$ of disjoint tiles such that $\bigcup_{t \in \cT} t = \cU$.
In particular, $\sum_{t \in \cT} \abs{t} = \abs{\cU} = 1$.
We use $A(\cT) \coloneqq \sum_{t \in \cT} \abs{A_t}$ to denote the area covered by choosing an area-maximal rectangle $A_t$ for each tile $t$ (the \emph{area covered by $\cT$}).

\paragraph{A Greedy Tile Packing Algorithm}
Let us revisit algorithm \TilePacking/ by \textcite{DBLP:journals/combinatorica/DumitrescuT15}.
\TilePacking/ processes the points from $P$ from top-right to bottom-left.
More formally, it orders $P = \set{p_1, p_2, \dots, p_n}$ such that $\dLine{p_i} \geq \dLine{p_{i+1}}$.
It then defines for each $p_i \in P$ the tile $t_i \coloneqq \set{q \in \cU | q \succeq p_i} \setminus \bigcup_{j=1}^{i-1} t_j$, yielding a tile packing $\cT = \set{t_1, t_2, \dots, t_n}$.
To build its solution to \LLARP/, \TilePacking/ picks for each $p \in P$ the rectangle $r_p$ as an (arbitrary) area-maximal rectangle $A_t \subseteq t$ in the tile $t$ containing $p$.
Thus, the total area covered by \TilePacking/ is $A(\cT)$.
\Cref{fig:tile_packing} illustrates the resulting tile packing.

Note that by this construction, the lower staircase points of each tile $t$ are input points.
Moreover, as already mentioned in \textcite{DBLP:journals/combinatorica/DumitrescuT15}, for each tile we can define a certain \emph{exclusive area} that does not contain an input point.
\begin{observation}%
\label{obs:exclusive_area:empty}
Consider the tile packing $\cT$ produced by \TilePacking/ for a set of input points $P$.
Fix a tile $t \in \cT$ and let $p \in P$ denote its anchor point.
Then the tile's \emph{exclusive area} $E_t \coloneqq \set{q \in \cU | \dLine{q} > \dLine{p} \land \exists q' \in \Gamma_t\colon q \prec q'}$ does not contain any input point from $P$.
\end{observation}
This \lcnamecref{obs:exclusive_area:empty} follows by noting that any such input point $p' \in E_t$ would be processed before $p$ by \TilePacking/ and \enquote{shield} at least one upper staircase point $q' \in \Gamma_t$ from $p$, preventing it from becoming an upper staircase point of tile $t$.

%% file: 30-general-approach.tex
\section{A General Approach for Lower Bounds}%
\label{sec:general_approach}

Here we present a general approach to derive lower bounds for the area covered by a given tile packing $\cT$.
Our approach relies on a suitable \emph{charging scheme} $\intoo{c_t}_{t \in \cT}$ that charges the area of each tile $t \in \cT$ to a \emph{charging area} $c_t > 0$.
We define $c^* \coloneqq \sum_{t \in \cT} c_t$ as the total charged area and $c_t / \abs{t}$ as the \emph{charging ratio} of tile $t$.

Assume we are given a piecewise differentiable\christoph{piecewise differentiable is not required, we only have to be able to evaluate $\xi$'s derivative at $\rho^*$}{} function $\xi\colon \intoc{0, 1} \to \R_{\geq 0}$ and a value $\rho^* \in \intoc{0, 1}$ with the following properties:
\begin{itemize}[nosep]
\item $\xi$ is point-convex in $\rho^*$ (see \cref{def:point_convex}) with $\xi'(\rho^*) < 0$,
\item $\xi(\rho^*)$ is an upper bound on the total charged area $c^*$, and
\item for any $t \in \cT$ the value $\xi(\rho_t)$ is a lower bound on $t$'s charging ratio $c_t / \abs{t}$.
\end{itemize}
The following \lcnamecref{lem:general_lower_bound} then uses $\xi$ to show that $\rho^*$ is a lower bound on the area covered by $\cT$.
\begin{lemma}%
\label{lem:general_lower_bound}
Consider a tile packing $\cT$ with a charging scheme $\intoo{c_t}_{t \in \cT}$ together with a function $\xi$ and a value $\rho^*$ as described above.
Then $\cT$ covers an area of at least $\rho^*$.
\end{lemma}
\begin{proof}
Since $\xi$ is point-convex in $\rho^*$, we get for the tangent $\tau(\rho) \coloneqq \xi(\rho^*) + \xi'(\rho^*) \cdot (\rho - \rho^*)$ of $\xi$ in $\rho^*$ that $\tau(\rho) \leq \xi(\rho)$ for all $\rho \in \intoc{0, 1}$.
Using $A(\cT) = \sum_{t \in \cT} \abs{A_t} = \sum_{t \in \cT} \abs{t} \cdot \rho_t$, the linearity of $\tau$ and the properties of $\xi$ we calculate
\begin{dmath}
\label{eqn:general_lower_bound:tangentbound}
     \tau\bigl(A(\cT)\bigr)
=    \tau\left( \sum_{t \in \cT} \abs{t} \cdot \rho_t \right)
=    \sum_{t \in \cT} \abs{t} \cdot \tau(\rho_t)
\leq \sum_{t \in \cT} \abs{t} \cdot \xi(\rho_t)
\leq \sum_{t \in \cT} \abs{t} \cdot \frac{c_t}{\abs{t}}
=    c^*
\leq \xi(\rho^*)
\end{dmath}.
On the other hand, by definition of $\tau$, we have
\begin{math}
  \tau\bigl(A(\cT)\bigr)
= \xi(\rho^*) + \xi'(\rho^*) \cdot \bigl(A(\cT) - \rho^*\bigr)
\end{math}.
Combining this with \cref{eqn:general_lower_bound:tangentbound} and rearranging yields
\begin{math}
     \xi'(\rho^*) \cdot A(\cT)
\leq \xi'(\rho^*) \cdot \rho^*
\end{math},
which yields the desired result after dividing by $\xi'(\rho^*) < 0$.
\end{proof}

%% file: 40-charging-scheme-and-weak-covering.tex
\section{Charging Scheme and Weak Covering Guarantee}%
\label{sec:charging_scheme+simple_covering_guarantee}

This \lcnamecref{sec:charging_scheme+simple_covering_guarantee} introduces the charging scheme we will use to derive our lower bounds for the area covered by the greedy algorithm \TilePacking/ (via the approach presented in \cref{sec:general_approach}).
Afterward we derive a first, simple\christoph{better:weak?}{} lower bound $\LBi\colon \intoc{0, 1} \to \R_{\geq 0}$ on the tiles' charging ratios (as described in \cref{sec:general_approach}) and prove that it has the properties necessary to apply \cref{lem:general_lower_bound}.
While comparatively simple, this already yields that \TilePacking/ covers at least a quarter of the unit square, almost tripling the original guarantee from~\cite{DBLP:journals/combinatorica/DumitrescuT15}.
\Cref{sec:strong_covering_guarantee} will refine $\LBi$ in order to derive our main result (\cref{thm:strong_covering_guarantee}).

\subsection{Charging Scheme}%
\label{sec:charging_scheme}

Given a tile packing $\cT$ from \TilePacking/, our charging scheme defines an area $C_t$ for each tile $t \in \cT$ and charges $t$'s area to $c_t \coloneqq \abs{C_t}$.
We first explain how $C_t$ is constructed from $t$.
Afterward, we prove useful properties about these areas and their relation to $\cT$.

\paragraph{Construction of $\bm{C_t}$}
Consider three points $p, q_1 = (x_1, y_1), q_2 = (x_2, y_2) \in \R^2$\christoph{move $p$ to the back of the list, otherwise one might confuse $p=(x_1,y_1)$.}{}  with $q_1, q_2 \succeq p$, $x_1 \leq x_2$, and $y_1 \geq y_2$.
Let $p' = (x_1, y_2)$.
The \emph{tower} $T_p(q_1, q_2)$ with \emph{base point} $p$ and \emph{peak} $p'$ is the rectangle formed by the lines $\dLine{p}$ (the tower's \emph{base}), $\aLine{q_1}$ (the tower's \emph{left side}), $\aLine{q_{i+1}}$ (the tower's \emph{right side}), and $\dLine{p_i}$ (the tower's \emph{top}).
If the subscript $p$ is omitted, the base point is assumed to be the origin $(0, 0)$.

To define a tile's charging area, fix a tile $t$ with anchor $p$ and $\Gamma_t = \set{q_1, q_2, \dots, q_k}$ ordered by increasing $x$-coordinate.
The charging area of tile $t$ is $C_t \coloneqq \bigcup_{i=1}^{k-1} T_p(q_i, q_{i+1})$.
We refer to $C_t$ as the \emph{crown} of tile $t$.
See \cref{fig:tile_crown} for an illustration of a tile's towers and crown.

\medskip

\begin{figure}
\vspace{-4ex}
\begin{minipage}[b]{0.45\textwidth-1em}
\begin{figure}[H]
\includegraphics{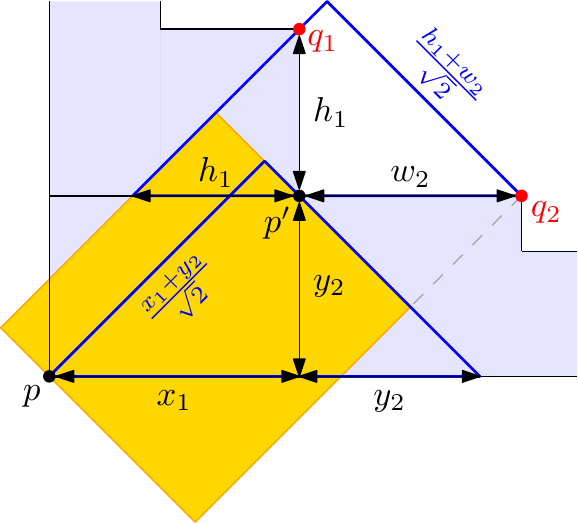}
\caption{
    $\abs{T_p(q_1, q_2)}$ is computed via the catheti of the blue triangles.
}\label{fig:tower_formula}
\end{figure}
\end{minipage}
\hfill
\begin{minipage}[b]{0.55\textwidth-1em}
\begin{figure}[H]
\centering
\includegraphics{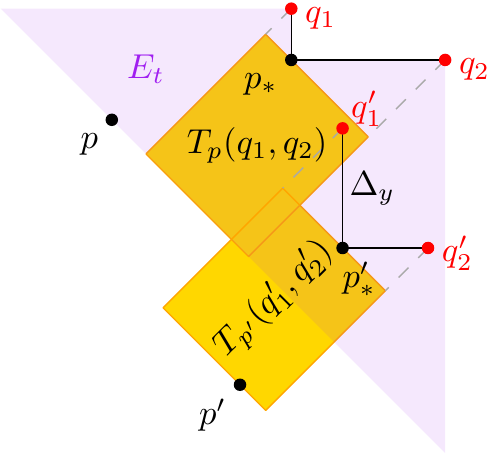}
\caption{
    Example for \cref{lem:no_crown_overlap}.
    The shown tower overlap has $p_*' \in E_t$, violating $t$'s exclusive area.
}\label{fig:crown_overlap}
\end{figure}
\end{minipage}
\end{figure}

The width and height of a tower $T_p(q_1, q_2)$ correspond to the side lengths of isosceles triangles (see \cref{fig:tower_formula}), which yields a formula for $\abs{T_p(q_1, q_2)}$.
By taking derivatives, we also get formulas for the change when moving $q_1$ or $q_2$ horizontally or vertically.
\begin{observation}%
\label{obs:tower_formula}
Consider $T_p(q_1, q_2)$ with $q_j - p = (x_j, y_j)$, $j \in \set{1, 2}$.
Let $w_2 \coloneqq x_2 - x_1$, and $h_1 \coloneqq y_1 - y_2$.
Then
\begin{math}
  \abs{T_p(q_1, q_2)}
= (x_1 + y_2) \cdot (w_2 + h_1) / 2
\end{math}.
\end{observation}

\begin{observation}%
\label{obs:upperstaircasemove:crown_change}
Consider $T_p(q_1, q_2)$ with $q_j - p = (x_j, y_j)$, $j \in \set{1, 2}$.
Let $w_2 \coloneqq x_2 - x_1$, and $h_1 \coloneqq y_1 - y_2$.
Fix $\alpha \in \R$ and consider the change of $\abs{T_p(q_1, q_2)}$ if either $q_1$ or $q_2$ are moved horizontally or vertically as a linear function of $\epsilon$:
\begin{alphaenumerate}[nosep]
\item
    If either $q_1(\epsilon) \coloneqq q_1 + (0, \alpha \cdot \epsilon)$ or $q_2(\epsilon) \coloneqq q_2 + (\alpha \cdot \epsilon, 0)$, then
    \begin{math}
      \partial \abs{T_p(q_1, q_2)}/\partial \epsilon
    = \alpha \cdot (x_1 + y_2) / 2
    \end{math}
    and
    \begin{math}
      \partial^2 \abs{T_p(q_1, q_2)}/\partial \epsilon^2
    = 0
    \end{math}.

\item
    If either $q_1(\epsilon) \coloneqq q_1 + (\alpha \cdot \epsilon, 0)$ or $q_2(\epsilon) \coloneqq q_2 + (0, \alpha \cdot \epsilon)$, then
    \begin{math}
      \partial \abs{T_p(q_1, q_2)}/\partial \epsilon
    = \alpha \cdot \bigl( w_2 + h_1 - (x_1 + y_2) \bigr) / 2
    \end{math}
    and
    \begin{math}
      \partial^2 \abs{T_p(q_1, q_2)}/\partial \epsilon^2
    = -\alpha^2
    \end{math}.
\end{alphaenumerate}
\end{observation}

\paragraph{Properties of the Charging Scheme}
The following results capture basic properties of our charging scheme.
First, we show that the defined charging areas are disjoint.
\begin{lemma}%
\label{lem:no_crown_overlap}
Consider the tile packing $\cT$ produced by algorithm \TilePacking/ for a set of input points $P$.
For any two different tiles $t, t' \in \cT$, we have $C_t \cap C_{t'} = \emptyset$.
\end{lemma}
\begin{proof}
Fix $t, t' \in \cT$ and let $p, p' \in P$ denote their respective anchors.
W.l.o.g., assume $\dLine{p} > \dLine{p'}$, such that \TilePacking/ processes $p$ before $p'$.
As crowns consist of towers, it is sufficient to show $T_p(q_1, q_2) \cap T_{p'}(q'_1, q'_2) = \emptyset$ for neighboring $q_1, q_2 \in \Gamma_t$ and $q'_1, q'_2 \in \Gamma_{t'}$.
Let $p_* \in \subseteq P$ and $p'_* \in P$ denote these towers' respective peak.
W.l.o.g., we assume $\aLine{p_*} < \aLine{p'_*}$ ($p_*$ lies left of $\aLine{p'_*}$); the other case follows symmetrically.

If $\dLine{p'_*} < \dLine{p}$, the towers are separated (the top of $T_{p'}(q'_1, q'_2)$ lies below the base of $T_p(q_1, q_2)$) and cannot intersect.
So assume $\dLine{p'_*} > \dLine{p}$.
Then we cannot have $p'_* \prec q_2$, since this would imply that $p'_*$ lies in the exclusive area of $t$, violating \cref{obs:exclusive_area:empty}.

Let $\Delta_y \coloneqq q'_1 - p'_*$ and note that $x(\Delta_y) = 0$.
Define $\tilde{q}_1 \coloneqq p_* - \Delta_y$ and note that $p'_* \not\succ \tilde{q}_1$, since otherwise $q'_1 = p'_* + \Delta_y \succ \tilde{q}_1 + \Delta_y = p_*$, which (together with $\dLine{p_*} > \dLine{p} > \dLine{p'}$) would mean that $p_*$ lies in the exclusive area of $t'$ (again violating \cref{obs:exclusive_area:empty}).

So $\aLine{p'_*} > \aLine{p_*}$, $p'_* \not\prec q_2$, and $p'_* \not\succ \tilde{q}_1$.
Together, these imply $x(p'_*) > x(q_2)$ and $y(p'_*) < y(\tilde{q}_1)$, which in turn imply $\aLine{p'_*} > \aLine{q_2 - \Delta_y}$ (see \cref{fig:crown_overlap}).
But then, the towers are separated, since
\begin{math}
  \aLine{q'_1}
= \aLine{p'_* + \Delta_y}
> \aLine{q_2 - \Delta_y + \Delta_y}
= \aLine{q_2}
\end{math}
($T_p(q_1, q_2)$'s right side lies left of $T_{p'}(q'_1, q'_2)$'s left side).
\end{proof}

\begin{figure}
\vspace{-2ex}
\begin{minipage}[b]{0.33\textwidth-1em}
\includegraphics{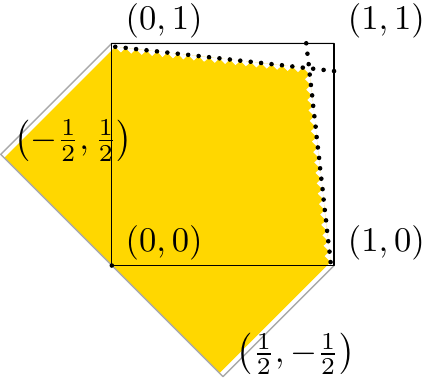}
\caption{
    Pentagon $\pentagon$ and point set $P_{\epsilon}$ from \cref{lem:crown_area_sum}.
}\label{fig:32crown}
\end{minipage}
\hfill
\begin{minipage}[b]{0.67\textwidth-1em}
\begin{figure}[H]
\centering
\includegraphics{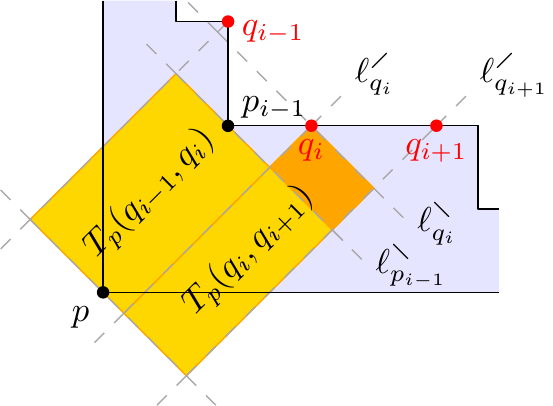}\vspace{-2ex}
\caption{
    Removing superfluous points from $\Gamma_t$ reduces $\abs{C_t}$ by the rectangle between $\aLine{q_i}$, $\aLine{q_{i+1}}$, $\dLine{q_i}$, and $\dLine{p_{i-1}}$.
}\label{fig:lt_point_removal}
\end{figure}
\end{minipage}
\end{figure}

The next \lcnamecref{lem:crown_area_sum}'s proof shows that all crowns lie inside a pentagon formed by $\cU$ and two isosceles triangles left and below of $\cU$ (see \cref{fig:32crown}).
With \cref{lem:no_crown_overlap} this implies that the total charging area is bounded by the pentagon's area.
\begin{lemma}%
\label{lem:crown_area_sum}
Consider the tile packing $\cT$ produced by algorithm \TilePacking/ for a set of input points $P$.
The total charging area of $\cT$ is $c^* \leq 3/2$.
Moreover, this bound is tight, since there are arbitrarily small $\epsilon > 0$ and input points $P_{\epsilon}$ for which $c^* \geq 3/2 - \epsilon$.
\end{lemma}
\begin{proof}
\newcommand*{\NW}{\mathrm{NW}}%
\newcommand*{\SE}{\mathrm{SE}}%
\newcommand*{\SW}{\mathrm{SW}}%
Define the points $\SW \coloneqq (0, 0)$, $\NW \coloneqq (0, 1)$, and $\SE \coloneqq (1, 0)$.
Let $\pentagon$ denote the pentagon enclosed by the lines $\dLine{\SW}$, $\aLine{\NW}$, $\aLine{\SE}$, $\hLine{\NW}$, and $\vLine{\SE}$ (see \cref{fig:32crown}).
Since $\abs{\pentagon} = 3/2$ and using \cref{lem:no_crown_overlap}, it is sufficient to show that $C_t \subseteq \pentagon$ for any $t \in \cT$.
For this, in turn, it is sufficient to show that any tower $T_p(q_1, q_2)$ of $C_t$ lies in $\pentagon$.

Fix such a tower $T_p(q_1, q_2)$.
Since $p \succ \SW$, we have $\dLine{p} \geq \dLine{\SW}$ (the base of $T_p(q_1, q_2)$ lies above the base of $\pentagon$).
Similarly, since $q_1, q_2 \in \overline{\cU} \subseteq \pentagon$, we have $\aLine{q_1} \geq \aLine{\mathrm{NW}}$ (the left side of $T_p(q_1, q_2)$ lies right of the left side of $\pentagon$) and $\aLine{q_2} \leq \aLine{\mathrm{SE}}$ (the right side of $T_p(q_1, q_2)$ lies left of the right side of $\pentagon$).
Finally, the topmost point $q_1 \in \overline{U}$ of $T_p(q_1, q_2)$ lies below $\hLine{NW}$ and the rightmost point $q_2 \in \overline{U}$ of $T_p(q_1, q_2)$ lies to the left of $\vLine{SE}$.
Together, we get $T_p(q_1, q_2) \subseteq \pentagon$.

For the tightness of the bound, choose $\epsilon > 0$ with $1 / \epsilon \in \N$.
Define $P_{\epsilon} = \set{\SW} \cup \set{(k \cdot \epsilon, 1 - k \cdot \epsilon^2), (1 - k \cdot \epsilon^2, k\epsilon) | k \in \set{1, 2, \dots, 1/\epsilon-1}}$.
As illustrated in \cref{fig:32crown}, the crown of tile $t$ with anchor $\SW$ converges towards $\pentagon$ as $\epsilon \to 0$, such that $\lim_{\epsilon \rightarrow 0} \abs{C_t} = 3/2$.
\end{proof}

\subsection{Weak Covering Guarantee for Greedy Tile Packings}%
\label{sec:weak_covering_guarantee}

This \lcnamecref{sec:weak_covering_guarantee} proves the following, slightly weaker version of \cref{thm:strong_covering_guarantee}:
\begin{theorem}%
\label{thm:weak_covering_guarantee}
For any input points, \TilePacking/ covers at least $25\%$ of the unit square.
\end{theorem}
Proving this not only serves as a warm-up to illustrate our approach before proving our main result but – as we will see in \cref{sec:strong_covering_guarantee} – brings us halfway towards proving \cref{thm:strong_covering_guarantee}.

So consider a tile packing $\cT$ produced by \TilePacking/ for some set of input points $P$.
To prove \cref{thm:weak_covering_guarantee} we follow the approach outlined in \cref{sec:general_approach}, using the charging scheme from \cref{sec:charging_scheme}.
That is, the area of $t \in \cT$ is charged to $c_t = \abs{C_t}$, where $C_t$ represents the crown of $t$.
To this end, define $\rho^* \coloneqq 1/4$ and the \emph{weak charging ratio bound}
\begin{dmath}
\label{eqn:LBi}
\LBi\colon \intoc{0, 1} \to \R_{\geq 0},\quad
\LBi(\rho) \coloneqq 2 \cdot (1 - \rho)
\end{dmath}.

As a linear function, $\LBi$ is trivially point-convex in $\rho^*$.
Moreover, $\LBi(\rho^*) = 3/2$ and thus, by \cref{lem:crown_area_sum}, $\LBi(\rho^*) \geq c^*$.
In the remainder of this \lcnamecref{sec:weak_covering_guarantee} we prove the following \lcnamecref{prop:LBi}, stating that $\LBi$ represents a lower bound on the charging ratio of any $t \in \cT$.
\begin{proposition}%
\label{prop:LBi}
For any tile $t$ we have $c_t / \abs{t} \geq \LBi(\rho_t)$.
\end{proposition}
Once this is proven, \cref{thm:weak_covering_guarantee} follows immediately by applying \cref{lem:general_lower_bound}.

\paragraph{A Lower Bound on the Charging Ratio}
To prove that $\LBi(\rho_t)$ lower bounds the charging ratio $c_t / \abs{t}$ of any tile $t \in \cT$, we gradually transform $t$ into a \enquote{simpler} tile $\tilde{t}$.
Our transformations ensure $\rho_{\tilde{t}} = \rho_t$ and $c_{\tilde{t}} / \abs{\tilde{t}} \leq c_t / \abs{t}$.
Eventually, $\tilde{t}$ will be simple enough to directly prove $c_{\tilde{t}} / \abs{\tilde{t}} \geq \LBi(\rho_{\tilde{t}})$.
The following notation expresses progress via such a transformation:
\begin{dmath}
\tilde{t} \preceq t
\quad\vcentcolon\Leftrightarrow\quad
\rho_{\tilde{t}} = \rho_t \text{ and } c_{\tilde{t}} / \abs{\tilde{t}} \leq c_t / \abs{t}
\end{dmath}.

As a simple example, note that both a tile's density and charging-ratio are invariant under translation and concentric scaling w.r.t.~its anchor.
This gives rise to the following transformation, which allows us to restrict our analysis to normalized tiles.
\begin{observation}%
\label{obs:transform:translate+scale}
For a tile $t$, let $\tilde{t}$ denote the translation of $t$ such that it is anchored in the origin and scaled by $1 / \abs{A_t}$ around the origin.
Then $\tilde{t} \preceq t$.
We call $\tilde{t}$ the \emph{normalization} of $t$.
\end{observation}

Consider a tile $t$ with anchor $p$.
A transformation may move one of $t$'s upper staircase points to the same $x$- or $y$-coordinate as another point from $\Gamma_t \cup \set{p}$, resulting in a degenerate tile with superfluous points in $\Gamma_t$ (see \cref{sec:preliminaries}).
The next \lcnamecref{lem:transform:degenerate} states that removing such superfluous points maintains an \enquote{equivalent} tile with a smaller crown.
\begin{lemma}%
\label{lem:transform:degenerate}
Consider a degenerate tile $t$.
The pruned tile $\tilde{t}$ with the same anchor but with $\Gamma_{\tilde{t}} \coloneqq \set{q \in \Gamma_t | \nexists q' \in \Gamma_t\colon q \preceq q'}$ covers the same points, is non-degenerate, and $c_{\tilde{t}} \leq c_t$.
\end{lemma}
\begin{proof}
Order $\Gamma_t = \set{q_1, q_2, \dots, q_k}$ by non-decreasing $x$-coordinate and let $q_0 = q_{k+1} = p$.
W.l.o.g.~assume there is some $i \in \set{1, \dots, k}$ with $y(q_i) = y(q_{i+1})$; the case of identical $x$-coordinates follows analogously.
Let $\tilde{t}$ denote the (possibly still degenerate) tile with anchor $p$ and $\Gamma_{\tilde{t}} = \Gamma_t \setminus \set{q_i}$.
Note that $\set{q \in \cU | q \succeq p \land q \prec q_i} \subseteq \set{q \in \cU | q \succeq p \land q \prec q_{i+1}}$, which implies $\tilde{t} = t$ and, thus, $\rho_t = \rho_{\tilde{t}}$.
Removing $q_i$ affects the towers $T_p(q_{i-1}, q_i)$ with peak $p_{i-1}$ and $T_p(q_i, q_{i+1})$ with peak $q_i$.
\Cref{fig:lt_point_removal} illustrates the situation.

We now show that $c_{\tilde{t}} \leq c_t$, such that $\tilde{t} \preceq t$; the \lcnamecref{lem:transform:degenerate}'s statement then follows by iteration.
If $i = k$, then $c_{\tilde{t}} = c_t - \abs{T(q_{i-1}, q_i)} \leq c_t$.
So assume $i < k$.
Then $c_{\tilde{t}} = c_t - \abs{\square} \leq c_t$, where $\square$ is the rectangle enclosed by the lines $\aLine{q_i}$, $\aLine{q_{i+1}}$, $\dLine{q_i}$, and $\dLine{p_{i-1}}$ (see \cref{fig:lt_point_removal}).
\end{proof}

\begin{figure}
\vspace{-2ex}
\begin{minipage}[b]{0.5\textwidth-1em}
\includegraphics{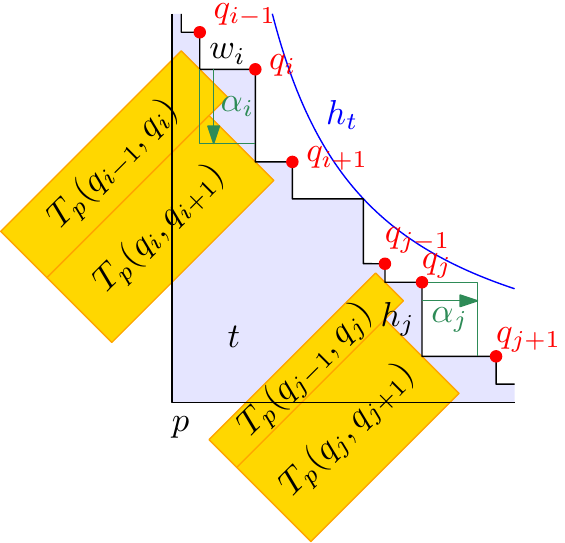}
\caption{Moving inner points in \Cref{lem:zwei-Punkte-Verschiebung}. Note that $\alpha_j>0$ and $\alpha_i<0$ in this case.}
\label{fig:zwei-Punkte-Verschiebung}
\end{minipage}
\hfill
\begin{minipage}[b]{0.5\textwidth-1em}
\begin{figure}[H]
\centering
\includegraphics{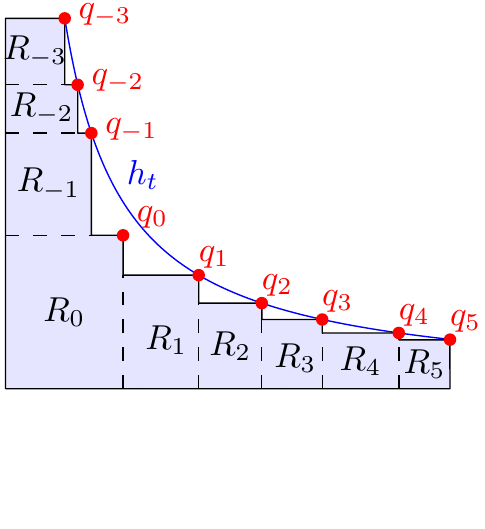}
\caption{Notation for \cref{prop:LBi} with $k=3$ and $m=5$.}
\label{fig:initial_ct_bound}
\end{figure}
\end{minipage}
\end{figure}

For the next transformation, remember that the hyperbola $h_t$ of a normalized tile $t$ contains exactly those upper staircase points $q \in \Gamma_t$ that form maximal rectangles in the tile.
We now prove that we can transform $t$ such that at most one $q \in \Gamma_t$ lies not on $h_t$.
\begin{lemma}%
\label{lem:zwei-Punkte-Verschiebung}
A normalized tile $t$ can be transformed into a tile $\tilde{t} \preceq t$ with $\abs{\Gamma_{\tilde{t}} \setminus h_{\tilde{t}}} \leq 1$.
\end{lemma}
\begin{proof}
Assume $\abs{\Gamma_t \setminus h_t} > 1$ and order $\Gamma_t = \set{q_1, q_2, \dots, q_k}$ by increasing $x$-coordinate.
To simplify border cases, define $q_0 = q_1$ and $q_{k+1} = q_k$.
Choose $q_i, q_j \in \Gamma_t \setminus h_t$ with $i < j$.
Consider the transformation $q_i(\epsilon) \coloneqq q_i + (0, \alpha_i \cdot \epsilon)$ and $q_j(\epsilon) \coloneqq q_j + (\alpha_j \cdot \epsilon, 0)$ with $\alpha_i, \alpha_j \in \R$.
Then the tile and crown areas become functions $t(\epsilon)$ and $c_t(\epsilon)$ of $\epsilon$.
We show that there are non-zero $\alpha_i, \alpha_j$ such that $t(\epsilon)$ and, thus, $\rho_t$ remain constant and $c_t(\epsilon)$ does not increase.
Eventually, this results in a tile $\tilde{t} \preceq t$ that has an additional point on $h_t$ or that has a degenerate staircase point which we can remove by \cref{lem:transform:degenerate}.
In both cases $\abs{\Gamma_t \setminus h_t}$ is reduced and the \lcnamecref{lem:zwei-Punkte-Verschiebung} follows by iteration.
\Cref{fig:zwei-Punkte-Verschiebung} illustrates the transformation.

The transformation changes only the towers $T(q_{i-1}, q_i)$, $T(q_i, q_{i+1})$, $T(q_{j-1}, q_j)$, and $T(q_j, q_{j+1})$.
For $l \in \set{0, 1, \dots, k+1}$ let $q_l = (x_l, y_l)$ and define $w_l \coloneqq x_l - x_{l-1}$ for $l \neq 0$ and $h_l \coloneqq y_l - y_{l+1}$ for $l \neq k+1$.
Then the transformation changes the tile according to
\begin{math}
  t'(\epsilon)
= \alpha_i \cdot w_i + \alpha_j \cdot h_j
\end{math}.
To keep $t$ constant we make this zero by setting $\alpha_j = -\alpha_i \cdot w_i / h_j$.

It remains to find a non-zero $\alpha_i$ such that $c_t(\epsilon)$ is non-increasing in $\epsilon$.
For this let $T_i(\epsilon) \coloneqq \abs{T(q_{i-1}, q_i)} + \abs{T(q_i, q_{i+1})}$ and define $T_j(\epsilon)$ analogously.
By \cref{obs:upperstaircasemove:crown_change},
\begin{math}
  T_l''(\epsilon)
= -\alpha_l^2
\end{math}
for $l \in \set{i, j}$.
This yields
\begin{math}
  c_t''(\epsilon)
= -\alpha_i^2 - \alpha_j^2
= -\alpha_i^2 \cdot (1 + w_i^2 / h_j^2)
< -\alpha_i^2
\end{math},
a negative constant.
This allows us to choose $\alpha_i \in \set{-1, +1}$ such that $c_t(\epsilon)$ is non-increasing.
\end{proof}

With these results, we are ready to prove our first covering guarantee for \TilePacking/.
\begin{proof}[Proof of \cref{prop:LBi}]
\newcommand*{\thigh}{t_{\text{high}}}
\newcommand*{\tlow}{t_{\text{low}}}
Consider an arbitrary tile $t$.
By \cref{obs:transform:translate+scale} and \cref{lem:zwei-Punkte-Verschiebung}, we can assume that $t$ is normalized and that $\abs{\Gamma_t \setminus h_t} \leq 1$.
If $\abs{\Gamma_t \setminus h_t} = 1$, let $q_0$ be that point, otherwise choose $q_0 \in \Gamma_t$ arbitrarily.
Order $\Gamma_t = \set{q_{-l}, \dots, q_k}$ by increasing $x$-coordinate.
To simplify border cases, define $q_{-l-1} = q_l$ and $q_{k+1} = q_k$.
To ease the notation, for $i \in \set{-l-1, \dots, k+1}$ we let $q_i = (x_i, y_i)$
We furthermore define $w_i \coloneqq x_i - x_{i-1}$ for $i \neq l-$ and $h_i \coloneqq y_i - y_{i+1}$ for $i \neq k+1$.
For $i \in \set{-l, \dots, k}$ inductively define $R_i \coloneqq \set{q \in t | q \prec q_i} \setminus \bigcup_{\abs{j} < i} R_j$.
Finally, for $i \in \set{1, \dots, k}$ let $T_i \coloneqq T(q_{i-1}, q_i)$ and for $i \in \set{-l, \dots, -1}$ let $T_i \coloneqq T(q_i, q_{i+1})$.
See \cref{fig:initial_ct_bound} for an illustration.

Note that $c_t = \sum_{i=-l}^{-1} \abs{T_i} + \sum_{i=1}^{k} \abs{T_i}$ and $\abs{t} = \sum_{j=-l}^{k} \abs{R_j}$.
We will first show that for $i \in \set{-l, \dots, k} \setminus \set{-1, 0, 1}$ we have $\abs{T_i} \geq 2 \abs{R_i}$.
Afterward, we show $\abs{T_{-1}} + \abs{T_1} \geq 2 \abs{R_{-1}}  + 2 \abs{R_{-1}} + 2 \abs{R_0} - 2$.
With these inequalities and since $\rho_t = \abs{A_t} / \abs{t} = 1/\abs{t}$ due to the normalization, the desired statement follows via
\begin{dmath}
\begin{aligned}
     c_t
=    \sum_{i=-l}^{-1} \abs{T_i} + \sum_{i=1}^{k} \abs{T_i}
\geq \sum_{i=-l}^{k} 2\abs{R_i} - 2
=    2 \abs{t} - 2
=    2 \abs{t} \cdot (1 - \rho_t)
=    \abs{t} \cdot \LBi(\rho_t)
.
\end{aligned}
\end{dmath}

We now show the above bounds, starting with $\abs{T_i} \geq 2 \abs{R_i}$ for $\abs{i} > 1$.
W.l.o.g.~we assume $i > 1$; the case $i < -1$ follows by symmetry.
Note that $i > 1$ implies $q_i, q_{i-1} \in h_t$ and thus (since $t$ is normalized) $y_j = 1/x_j$ for $j \in \set{i-1, i}$.
This yields
\begin{math}
x_i / y_{i-1} = x_{i-1} \cdot x_i
\end{math}
as well as
\begin{math}
w_i / h_{i-1} = (x_i - x_{i-1}) / (y_{i-1} - y_i) = x_{i-1} \cdot x_i
\end{math}.
We use these identities together with $\abs{R_i} = w_i \cdot y_i$ to bound the formula for $\abs{T_i}$ from \cref{obs:tower_formula}:
\begin{dmath}
\begin{aligned}
\abs{T_i}
&=
\frac{1}{2} \cdot (x_{i-1} + y_i) \cdot (w_i + h_{i-1})
=
w_i \cdot y_i \cdot \frac{1}{2} \cdot \left( 1 + \frac{x_{i-1}}{y_i} \right) \cdot \left( 1 + \frac{h_{i-1}}{w_i} \right)
\\&=
\abs{R_i}     \cdot \frac{1}{2} \cdot (1 + x_{i-1} \cdot x_i) \cdot \left( 1 + \frac{1}{x_{i-1} \cdot x_i} \right)
=
\abs{R_i}     \cdot \frac{1}{2} \cdot \frac{{(1 + x_{i-1} \cdot x_i)}^2}{x_{i-1} \cdot x_i}
\geq 2 \abs{R_i}
,
\end{aligned}
\end{dmath}
where the inequality follows since $x \mapsto {(1 + x)}^2 / x$ takes its minimum over $\intco{0, \infty}$ at $x = 1$.

It remains to show that $\abs{T_{-1}} + \abs{T_1} \geq 2 \abs{R_{-1}}  + 2 \abs{R_{-1}} + 2 \abs{R_0} - 2$.
Note that by our definition $q_{k+1} = q_k$, if $k = 0$ we have $\abs{R_1} = 0$ and $\abs{T_1} = 0$.
Similarly, if $l = 0$ then $\abs{R_{-1}} = 0$ and $\abs{T_{-1}} = 0$.
We assume that not both $k$ and $l$ are zero, as otherwise $\LBi(\rho_t) = \LBi(1) = 0$ and the \lcnamecref{prop:LBi} becomes trivial.
W.l.o.g.~let $k > 0$; the other case follows symmetrically.

For $\alpha \in \set{-1, +1}$ (which we fix later) and $\epsilon \geq 0$ define the transformation $y_0(\epsilon) \coloneqq y_0 + \alpha \cdot \epsilon \in \intcc{y_1, 1/x_0}$\footnote{
     These boundaries ensure that the tile remains valid and normalized.
     Note that if $l = 0$, moving $q_0$ upward also causes the dummy point $q_{-1}$ to move upward, such that $\abs{R_{-1}}$ and $\abs{T_{-1}}$ remain zero.
}, which moves $q_0$ either up- or downward, depending on $\alpha$.
Thus, with $f(\epsilon) \coloneqq \abs{T_{-1}} + \abs{T_1} - 2 \abs{R_{-1}} - 2 \abs{R_1} - 2 \abs{R_0} + 2$ our goal becomes to prove $f(0) \geq 0$.
To this end, consider how $f(\epsilon)$ changes with $\epsilon$.
The rectangles $\abs{R_j}$ ($j \in \set{-1, 0, 1}$) change linearly or remain constant.
By \cref{obs:upperstaircasemove:crown_change}, $\partial^2\abs{T_1} / \partial\epsilon^2 = 0$.
Similarly, if $l > 0$ we have $\partial^2\abs{T_{-1}} / \partial\epsilon^2 = -\alpha^2 = -1$ by \cref{obs:upperstaircasemove:crown_change}, and if $l = 0$ we have $\partial^2\abs{T_{-1}} / \partial\epsilon^2 = 0$ (because $\abs{T_{-1}}$ remains zero).
Thus, in all cases
\begin{math}
     f''(\epsilon)
\leq 0
\end{math}.
Then for one of the choices $\alpha \in \set{-1, +1}$ the function $f$ must be non-increasing, meaning its minimum $f_{\min}$ lies at one of the borders, where either $y_0 = y_1$ or $y_0 = 1/x_0$.
We consider both possibilities and show that each time $f_{\min} \geq 0$ (which finishes the proof, since $f(0) \geq f_{\min} \geq 0$).

If in $f_{\min}$ we have $y_0 = 1/x_0$, let $\thigh$ denote the corresponding tile.
Note that $q_0$ lies on the hyperbola $h_t$.
But then $\abs{R_0} = 1$ and, thus, $f_{\min} = \abs{T_{-1}} + \abs{T_1} - 2 \abs{R_{-1}} - 2 \abs{R_1}$.
Moreover, with $q_0 \in h_t$ we can apply the calculations 
for $\abs{i} > 1$ to get $\abs{T_{-1}} \geq 2 \abs{R_{-1}}$ and $\abs{T_1} \geq 2 \abs{R_1}$, such that $f_{\min} \geq 0$.

So assume that in $f_{\min}$ we have $y_0 = y_1$ and let $\tlow$ denote the corresponding tile.
Note that $R_0$ and $R_1$ form a rectangle from the origin to the point $q_1$ on $h_t$, such that $\abs{R_0} + \abs{R_1} = 1$.
Thus, $f_{\min} = \abs{T_{-1}} + \abs{T_1} - 2 \abs{R_{-1}}$.
Define the (degenerate) tile $t'$ with $\Gamma_{t'} = \set{q_{-1}, q_0, q_1}$ and anchor $p$, such that its crown area is $c_{t'} = \abs{T_1} + \abs{T_2}$.
By \cref{lem:transform:degenerate}, for the (non-degenerate) tile $\tilde{t}'$ with $\Gamma_{\tilde{t}'} = \Gamma_{t'} \setminus \set{q_0}$ we have $c_{\tilde{t}'} \leq c_{t'}$.
The crown $c_{\tilde{t}'}$ consists of the single tower $T(q_{-1}, q_1)$.
Since $q_{-1}, q_1 \in h_t$, we can apply the calculations 
for $\abs{i} > 1$ to get $c_{\tilde{t}'} = \abs{T(q_{-1}, q_1)} \geq 2 \abs{R_{-1}}$.
Putting everything together we get
\begin{equation*}
     f_{\min}
=    \abs{T_{-1}} + \abs{T_1} - 2 \abs{R_{-1}}
=    c_{t'} - 2 \abs{R_{-1}}
\geq c_{\tilde{t}'} - 2 \abs{R_{-1}}
\geq 0 . \qedhere
\end{equation*}
\end{proof}

%% file: 50-strong-covering.tex
\section{Strong Covering Guarantee for Greedy Tile Packings}%
\label{sec:strong_covering_guarantee}

This \lcnamecref{sec:strong_covering_guarantee} proves our strong covering guarantee for \TilePacking/, namely \cref{thm:strong_covering_guarantee}.
We use the same approach as for our weak covering guarantee from \cref{sec:weak_covering_guarantee} but derive a stronger bound on the tiles' charging ratios.
More exactly, instead of $\LBi$ we use
\begin{dmath}
\label{eqn:LBii}
\LBii\colon \intoc{0, 1} \to \R_{\geq 0},\quad
\LBii(\rho) \coloneqq
\begin{cases}
1 - \rho \cdot \bigl( 1 + \sinh(1 - 1/\rho) \bigr) & \text{, if $\rho \leq 1/2$} \\
\LBi(\rho) = 2 \cdot (1 - \rho)                    & \text{, if $\rho > 1/2$.}
\end{cases}
\end{dmath}

Most properties required for our approach from \cref{sec:general_approach} are easily verified for $\LBii$ (whose function graph can be seen in \cref{fig:xi_graph} in \cref{app:auxiliary}).
Indeed, for $\rho^* \coloneqq \LBii^{-1}(3/2) \approx 0.3901$, we have $\LBii'(\rho^*) \approx -5.1 < 0$.
Moreover, $\LBii$ is point-convex in $\rho^*$, since it is convex on $\intoc{0, 1/2}$ and on $\intoc{1/2, 1}$ its tangent $t_{\xi}(\rho)$ in $\rho^*$ lies below $\LBii$ ($t_{\xi}$ is steeper and $t_{\xi}(1/2) \approx 0.94 < 1 = \LBi(1/2)$).
Also, by choice of $\rho^*$ and by \cref{lem:crown_area_sum}, we have $\LBii(\rho^*) = 3/2 \geq c^*$ for the total charged area $c^*$ of a tile packing $\cT$ produced by \TilePacking/.

The following \lcnamecref{prop:LBii} states the remaining required property from \cref{sec:general_approach}.
\begin{proposition}%
\label{prop:LBii}
For any tile $t$ we have $c_t / \abs{t} \geq \LBii(\rho_t)$ and this bound is tight.
\end{proposition}
With this, \cref{thm:strong_covering_guarantee} follows by applying \cref{lem:general_lower_bound}.
The remainder of this section outlines the analysis of this \lcnamecref{prop:LBii}.
Full formal statements and proofs are given in \cref{subsection:improving_bound}.

\begin{figure}\vspace{-2ex}
\begin{minipage}[b]{120pt}
\begin{figure}[H]
\makebox[120pt][r]{\includegraphics{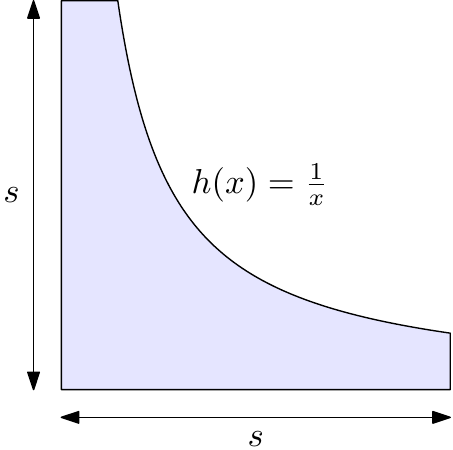}}
\caption{Low-density tile $\wcTileLD{s}$ for which $\LBii$ is tight.}
\label{fig:extreme_case_low_density}
\end{figure}
\end{minipage}
\hfill
\begin{minipage}[b]{120pt}
\begin{figure}[H]
\makebox[120pt][r]{\includegraphics{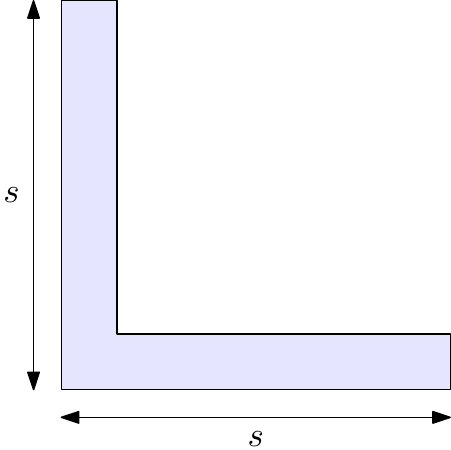}}
\caption{High-density tile $\wcTileHD{s}$ for which $\LBii$ is tight.}
\label{fig:extreme_case_high_density}
\end{figure}
\end{minipage}
\hfill
\begin{minipage}[b]{120pt}
\begin{figure}[H]
\makebox[120pt][r]{\includegraphics{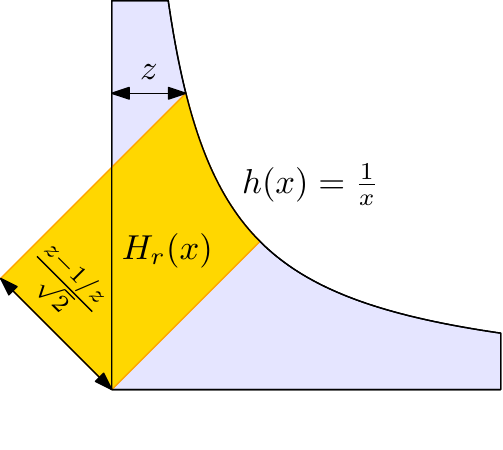}}
\caption{Crown contribution of a slide.}
\label{fig:Hrformula}
\end{figure}
\end{minipage}
\end{figure}

\paragraph{Transformation to Worst-case Tiles}
For tiles of density larger than $1/2$, \cref{prop:LBii} follows from \cref{prop:LBi}, since in this regime $\LBii(\rho) = \LBi(\rho)$.
The tightness for such high densities follows since for any $\rho \in \intoc{1/2, 1}$ there is a (symmetric) \emph{step tile} $t = \wcTileHD{s}$ of size $s$ (depicted in \cref{fig:extreme_case_high_density}) with density $\rho_t = \rho$ and $c_t / \abs{t} = \LBii(\rho_t)$.
Thus, we restrict our further study to tiles of density at most $1/2$.
We will show how to gradually transform any such tile $t$ into a (symmetric) \emph{hyperbola tile} $\wcTileLD{s} \preceq t$ of size $s$ (depicted in \cref{fig:extreme_case_low_density}).
Again, the tightness follows since for any $\rho \in \intoc{0, 1/2}$, we can choose $s$ such that tile $t = \wcTileLD{s}$ has density $\rho_t = \rho$ and $c_t / \abs{t} = \LBii(\rho_t)$.

Before we outline the transformation process into such worst-case low-density tiles, we need to cope with the fact that $\wcTileLD{s}$ is not a staircase polygon and, thus, not captured by our tile definition.
However, one can see $\wcTileLD{s}$ as the result of defining $\Gamma_t$ as $k$ equally spaced points from the hyperbola $\set{(x, y) \in \intco{0, s} | y = 1/x}$ and taking the limit $k \to \infty$.
The next paragraph formalizes this intuition by introducing \emph{generalized tiles} and some related notions.

\paragraph{Generalized Tiles}
As normal tiles, a \emph{generalized tile} $t \subseteq \cU$ is defined by its anchor $p \in \cU$ and $\Gamma_t \subseteq \overline{\cU}$ with $q \preceq p$ and $q \not\preceq q'$ for all $q \neq q'$ from $\Gamma_t$.
The only difference is that $\Gamma_t$ may be infinite.
All other tile definitions (e.g., point set $t$, rectangle $A_t$, or density $\rho_t$) stay intact.

From now on the term \emph{tile} always refers to a generalized tile.
We require that the $x$-coordinates of $\Gamma_t$ can be partitioned into $k$ inclusion-wise maximal, closed intervals $I_1, I_2, \dots, I_k$, ordered by increasing $x$-coordinates.
For $i \in \set{1, 2, \dots, k}$ let $q_i^-, q_i^+ \in \Gamma_t$ denote the points realizing the left- and rightmost $x$-coordinate of $I_i$, respectively.
Note that $I_i$ may be a point interval, such that $q_i^- = q_i^+$.
A \emph{section} of $\Gamma_t$ is a tuple as follows:
\begin{itemize}[nosep]
\item a \emph{step} $(q_i^+, q_{i+1}^-)$, if $q_i^+, q_{i+1}^- \in h_t$;
\item a \emph{slide} $(q_i^-, q_i^+)$, if $q_i^- \neq q_i^+$ and $\set{q \in \Gamma_t | x(q) \in I_i} \subseteq h_t$;
\item a \emph{double step} $(q_{i-1}^+, q_i^-, q_{i+1}^-)$, if $q_{i-1}^+, q_{i+1}^- \in h_t$ and $q_i^- = q_i^+ \notin h_t$; or
\item the \emph{corners} $(q_1, q_2)$, if $q_1 \notin h_t$ and $q_2 \in h_t$ as well as $(q_{k-1}, q_k)$ if $q_k \notin h_t$ and $q_{k-1} \in h_t$.
\end{itemize}
After applying \cref{lem:zwei-Punkte-Verschiebung}, all tiles resulting from our transformations can be described as a sequence of such sections.
\Cref{fig:generalized_tile} illustrates generalized tiles and the different sections.

\paragraph{Crown Contribution of Slides}
Our charging scheme from \cref{sec:charging_scheme} generalizes naturally to slides by considering them as the limit of $k \to \infty$ equally spaced upper staircase points.
This yields the following complement to \cref{obs:tower_formula}.
\begin{observation}%
\label{obs:hyperbola_formula}
For a tile $t$, the contribution of a slide $(q_1,q_2)$ to the crown $C_t$ is $H(q_1,q_2):=\left[\ln(z)+1/4\cdot (z^2-z^{-2})\right]_{x_1}^{x_2}$.
\end{observation}
\begin{proof}
Rotate the hyperbola by $\pi/4$ around the bottom-left corner of $t$, to obtain the rotated hyperbola $h_r(x)=\sqrt{x^2+2}$.
The contribution can then be calculated by integration:
The indefinite integral under $h_r(x)$ is
\begin{math}
          H_r(x)
\coloneqq \int {h_r(x)}dx
=         x/2\cdot \sqrt{2+x^2}+\arcsinh{(x/\sqrt{2})}
\end{math},
and for $x = (z-1/z)/\sqrt{2}$, we get $H_r((z-1/z)/\sqrt{2}) = \ln(z) + 1/4 \cdot (z^2-z^{-2})$ (see \cref{fig:Hrformula}).
\end{proof}

\paragraph{Overview of the Transformation Process}
\Cref{fig:transformation_plan} gives an overview of how we gradually transform an arbitrary tile $t$ with density $\rho_t \leq 1/2$ into a worst-case hyperbola tile $\wcTileLD{s}$.
Starting with an arbitrary tile $t$ (\rom{1}), \cref{lem:slope_lemma} either enforces $\Gamma_t \subseteq h_t$ or exactly one $q \notin h_t$, can be forced to be in a double-step $(q_1, q, q_2)$ with $x(q_1) \leq 1 \leq x(q_2)$ (\rom{2}), or to be in corner $(q, q')$ with $x(q') \geq 1$ or corner $(q', q)$ with $x(q') \leq 1$ (\rom{3}).

The next transformation from (\rom{2})/(\rom{3}) to (\rom{4})/(\rom{5}) is based on smaller modifications (\cref{lem:outer_step, lem:shorter_side, lem:step_merging}) which are combined in \cref{lem:enforced_high_density} to eliminate all but one step (while increasing, e.g., slides).
\Cref{lem:outer_step} \enquote{squeezes} the tile to broaden very thin left- or rightmost steps $(q, q')$.
\Cref{lem:shorter_side} moves steps with an adjacent slide along the hyperbola towards either the left or bottom of the tile, whichever is nearer.
\Cref{lem:step_merging} transforms two consecutive steps left of $(1, 1)$ into a step and a slide (or symmetrical on the right side of $(1, 1)$).

After we are down to only one step, \cref{lem:inner_nonhyp_point_or_hyperbola, lem:hyp_merge_with_boundary_nh_step} show for (\rom{4}) and (\rom{5}), respectively, that either $\Gamma_t \subseteq h_t$ or $t$ contains no slides which (together with \cref{lem:hilfslemma}) reduces the remaining cases to those illustrated in (\rom{6}) to (\rom{9}).
For each of these we separately show $\wcTileLD{s}\preceq t$ (\cref{lem:step_plus_hyperbola_section, lem:double_step, lem:step_plus_double_step, lem:step_plus_boundary_nh_step}), proving \cref{prop:LBii}.

%% file: 60-upper-bound.tex
\section{Upper Bound} \label{sec:upper-bound}
\noindent
\begin{minipage}{\textwidth-131.634pt-2em}
\setlength{\parindent}{15pt}
\noindent To show \cref{thm:worst_case_covering}, we construct a point set where \TilePacking/ fails to cover an area larger than roughly $(1-e^{-2})/2$.
Our goal is to construct a tile $\hat{t}$ at the origin where each maximal rectangle has the same size $A$.
We therefore place a large number of $k$ points $q_i$ densely on a hyperbola $h_A$ centered in the origin.
The remaining tiles are constructed such that they have a density close to $1/2$.
To this end we add for each point $q_i$ on $h_A$ a set of (almost) evenly spaced points $p_{i,j}$ with distance roughly $\varepsilon$ between each other, see \cref{fig:upper_bound_simplified}.
\end{minipage}
\hfill
\begin{minipage}{131.634pt}
\begin{figure}[H]
\centering
\vspace{-10ex}
\includegraphics{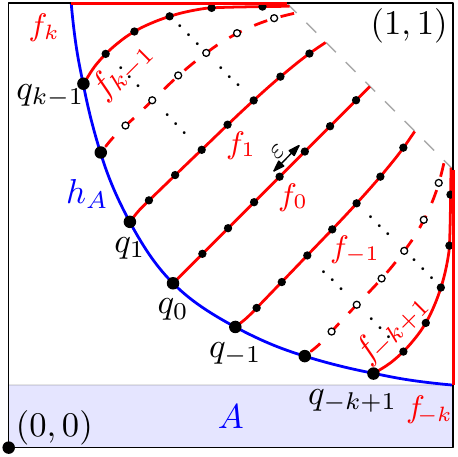}
\caption{Our construction.}
\label{fig:upper_bound_simplified}
\end{figure}
\end{minipage}\smallskip

To get a density close to $1/2$, the exact coordinates of the points $p_{i,j}$ must be chosen carefully: we place the points on arcs of functions $f_i$ described by differential equations, where each $f_i$ depends on the two neighboring curves $f_{i-1}$ and $f_{i+1}$.
Formal definitions of the point set $P_{\varepsilon,k}$ \mbox{and the functions $f_i$ are given in \cref{sec:full-proof-upper-bound}}.

We now sketch the proof of \cref{thm:worst_case_covering}.
For simplicity, assume that \TilePacking/ on $P_{\varepsilon,k}$ covers half of $\hat{\cU} = \cU \setminus \hat{t}$ and an area of $A$ of $\hat{t}$.
Both assumptions introduce an error; in \cref{sec:full-proof-upper-bound} we show that both error terms vanish as $k$ goes to infinity and $\varepsilon$ goes to zero.
In the full proof, we first show that the $f_i$ are well-behaved: their arcs do not intersect each other, and they only intersect $h_A$ at $q_i$.
This allow us to show the aforementioned density close to $1/2$ for all tiles $\neq \hat{t}$ (unless they are too close to $(1,1)$, in which case they have negligible area.)
The upper bound then follows by optimizing the parameter $A$.

\enlargethispage{2\baselineskip}
\begin{proof}[Proof Sketch for \cref{thm:worst_case_covering}]
We analyze the area $\rho$ covered by \TilePacking/ on $P_{\varepsilon,k}$.
Let $\hat{\cU} = \cU \setminus \hat{t}$.
Assume that \TilePacking/ covers half of $\hat{\cU}$ and an area of $A$ of $\hat{t}$.
Then
\begin{math}
\displaystyle
\rho \leq A+|\hat{\cU}|/2 =  A+(1-|\hat{t}|)/2 \leq A + \bigl( 1-(A+\int_{A}^{1}{A/x\ \mathrm{d} x}) \bigr)/2 = (1+A+A\ln{A})/2
\end{math}.
Minimizing this term leads to $\rho=(1-e^{-2})/2$ at $A=e^{-2}$.\christoph{$\le$'s should be $=$ since we are assuming exact half-coverage in $\hat{U}$ and exactly $A$ coverage in $\hat{t}$}{}
\end{proof}

%% file: 90-appendix.tex
\clearpage
\section{Auxiliary Notions and Results}%
\label{app:auxiliary}
\makeatletter
\@mkboth{APPENDIX}{APPENDIX}
\makeatother
\christoph{move appendix header above figure?}{}
The next \lcnamecref{lem:general_point_sets} proves that when analyzing algorithm \TilePacking/ we can, w.l.o.g., restrict ourselves to input point sets $P$ that are in general position.
\begin{lemma}%
\label{lem:general_point_sets}
\textsc{TilePacking} achieves the same bound on point sets in general position as for arbitrary point sets.
\end{lemma}
\begin{proof}
Assume that \textsc{TilePacking} chooses the points in some order $p_1,\dots,p_n$ while producing a tile packing $\cT$.
Move all points except $p_n=(0,0)$ to the top right by at most some small $\varepsilon > 0$ such that the processing order by \textsc{TilePacking} stays the same and the point set is in general position. \textsc{TilePacking} will produce a new tile packing $\cT'$.
The upper staircase points of each tile $t$ move by at most $\varepsilon$ in either direction and thus $A_t$ may increase by at most
$\LDAUOmicron{\varepsilon}$.
Hence $A(\cT')\le A(\cT)+\abs{P} \LDAUOmicron{\varepsilon}$, which proves the statement for $\varepsilon\rightarrow 0$.
\end{proof}

The following \lcnamecref{def:point_convex} captures a \enquote{local} version of convexity.
We use in \cref{sec:general_approach} to derive a lower bound on the area covered by a tile packing.
\begin{definition}%
\label{def:point_convex}
A function $f\colon I \to \R$ is said to be \emph{point-convex} in $x \in I$ if $f$ is differentiable in $x$ and the tangent $t$ of $f$ in $x$ satisfies $t(x) \leq f(x)$ for all $x \in I$.
\end{definition}
An immediate consequence of \cref{def:point_convex} is that Jensen's inequality holds for $f$ in $x$.
That is, if $f$ is point-convex in $x$ and $x = \sum_{i=1}^{k} \gamma_i \cdot x_i$ is a convex combination of $x_i \in I$ (i.e., $\gamma_i \geq 0$ and $\sum_{i=1}^{k} \gamma_i = 1$), then
\begin{dmath}
     f(x)
=    t(x)
=    \sum_{i=1}^{k} \gamma_i \cdot t(x_i)
\leq \sum_{i=1}^{k} \gamma_i \cdot f(x_i)
\end{dmath}.

\begin{figure}[b!]
\begin{minipage}[b]{0.5\textwidth-1em}
\begin{figure}[H]
\centering
\includegraphics[width=\linewidth]{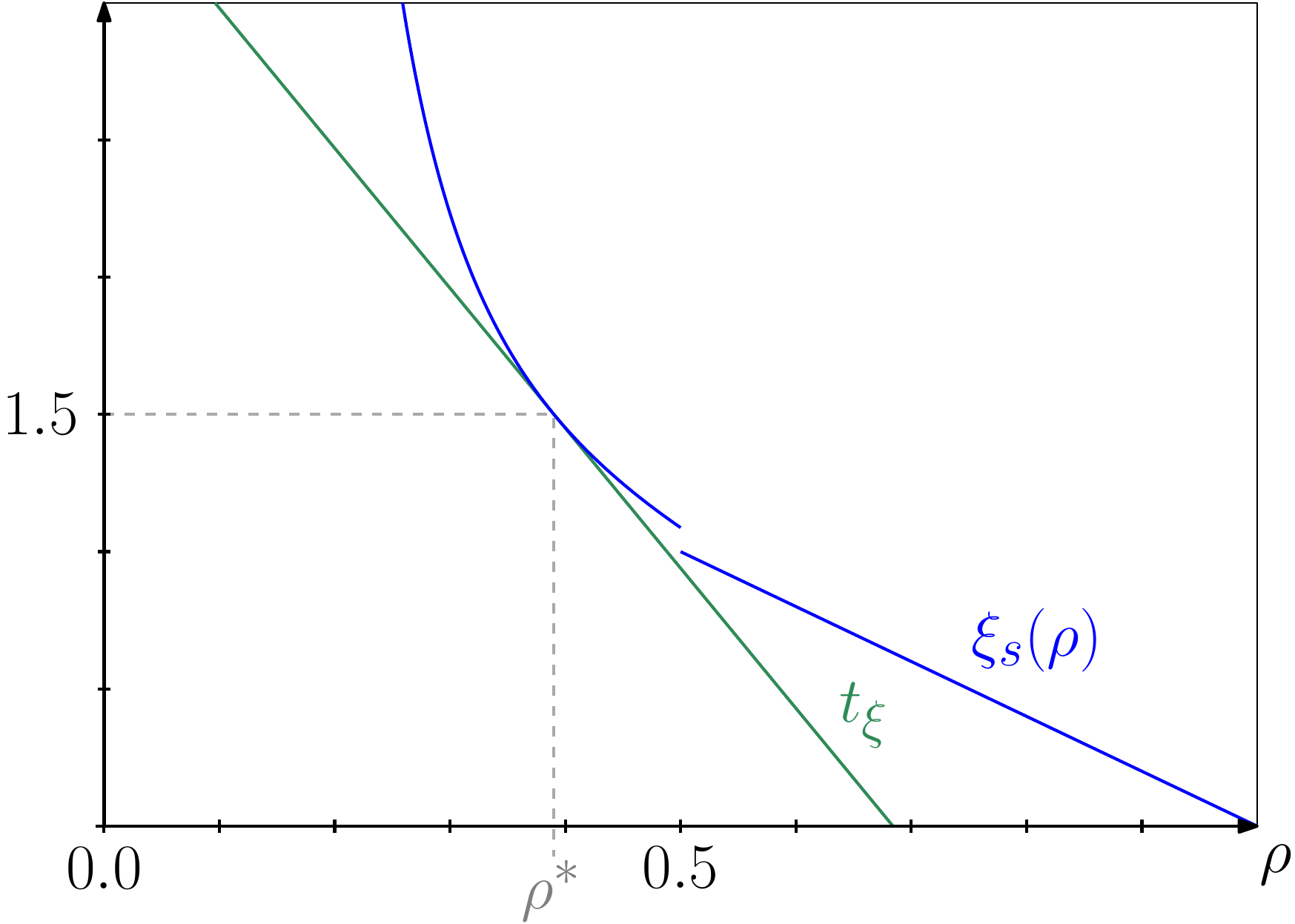}
\caption{
    The plot shows our strong lower bound function $\LBii$ from \cref{sec:strong_covering_guarantee}. The green line is its tangent $t_{\xi}$ in $\rho^* = \LBii^{-1}(3/2)$. Note that this illustrates that $\LBii$ is point-convex in $\rho^*$.
    \christoph{inconsistency: $\rho$ argument is only shown at $\LBii$, but not in $t_{\xi}$.}{}
}\label{fig:xi_graph}
\end{figure}
\end{minipage}
\hfill
\begin{minipage}[b]{0.5\textwidth-1em}
\begin{figure}[H]
\centering
\includegraphics[width=\linewidth]{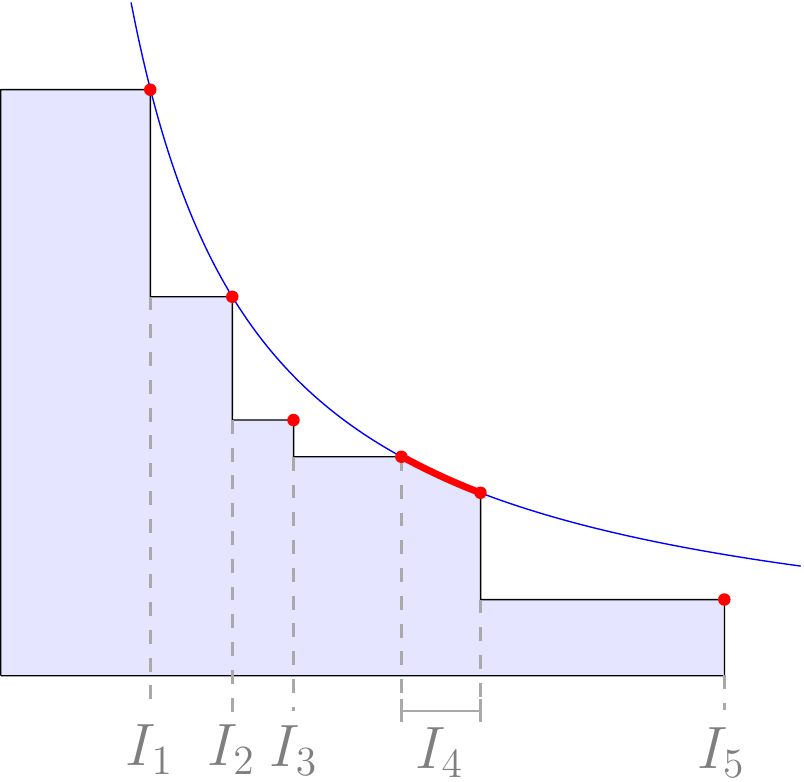}
\caption{
    A generalized tile with four sections formed by the five intervals $I_1$ to $I_5$ (of which only $I_4$ is a proper interval).
    They form a step (between $I_1$ and $I_2$), a double-step (between $I_2$ and $I_4$), a slide ($I_4$), and a corner ($I_4, I_5$).
}\label{fig:generalized_tile}
\end{figure}
\end{minipage}
\end{figure}

\section{Details for the Improved Bound}
\label{subsection:improving_bound}

In this section we show the omitted details for the strong covering guarantee of \Cref{sec:strong_covering_guarantee}. Note that we will use $[F(x)]_a^b:=F(b)-F(a)$ as a shorthand notation. Additionally we will use $\vec{x}\rightarrow \vec{x}'$ as a notation for substituting $\vec{x}$ with $\vec{x}'$. Due to \Cref{lem:zwei-Punkte-Verschiebung} we will always assume that $t$ satisfies $|\Gamma_t\setminus h_t|\leq1$.\christoph{notation is missing meaning of  $\left.term\right|_{a\rightarrow b}$. The $a\rightarrow b$ isnt consistent either, sometimes we use $a=b$ and such.}{}

\begin{remark}
    All transformation Lemmas obviously hold up to mirroring at the $(x=y)$-axis.
\end{remark}

\begin{observation}
\label{obs:transformation}
In the following we will make often use of very similar arguments to rule out specific structures of tiles. For this argument we define the area $|t(\vec{x})|$ and crown $C(\vec{x})$ of a tile as a function of some (general) parameters. Usually these will be the $x_i=x(q_i)\in\ammaG_t$, but we leave them as general parameters for now. We then introduce a transformation on these parameters, i.e., $\vec{x}(\varepsilon)$ with $\vec{x}(0)=\vec{x}$ as well as defining $\vec{X}=\vec{x}'|_{\varepsilon=0}$ and $\vec{Y}=\vec{x}''|_{\varepsilon=0}$. As our transformation needs to leave the area invariant we will enforce that $|t(\vec{x}(\varepsilon))|=|t(\vec{x})|$ for all $\varepsilon$.

We then obtain that $\partial_\varepsilon \frac{C(\vec{x}(\varepsilon))}{|t(\vec{x}(\varepsilon))|}|_{\varepsilon=0}= \frac{\partial_\varepsilon C(\vec{x}(\varepsilon))|_{\varepsilon=0}}{|t(\vec{x})|}=\frac{\vec{X}\vec{\nabla}C(\vec{x})}{|t(\vec{x})|}$

Assuming that all parameters can be increased or decreased, the only way we cannot improve the ratio with a transformation is when:

\begin{equation}
\label{eqn:transformation1}
    \vec{X}\vec{\nabla}C(\vec{x})=0
\end{equation}

As it is still possible that our transformation is in a local maximum of, we therefore also analyze the second derivative:

\begin{align*}
    &\partial^2_\varepsilon \frac{C(\vec{x}(\varepsilon))}{|t(\vec{x}(\varepsilon))|}|_{\varepsilon=0}\\
    &= \frac{\partial^2_\varepsilon C(\vec{x}(\varepsilon))|_{\varepsilon=0}}{|t(\vec{x})|}\\
    &= \frac{\partial_\varepsilon x'_j(\varepsilon) C_j(\vec{x}(\varepsilon))}{|t(\vec{x})|}|_{\varepsilon=0}\\
    &= \frac{x''_j(\varepsilon) C_j(\vec{x}(\varepsilon))+x'_j(\varepsilon) x'_k(\varepsilon) C_{jk}(\vec{x}(\varepsilon))}{|t(\vec{x})|}|_{\varepsilon=0}=\frac{\vec{Y}\vec{\nabla}C(\vec{x})+\vec{X}H_{C}\vec{X}}{|t(\vec{x})|}
\end{align*}

Where $C_j$ is shorthand for $\partial_{x_j}C(x_i)$ and $H_C$ is the Hessian of $C(x_i)$.
Thus, we have a local minimum and can thus not improve the ratio, as long as:

\begin{equation}
\label{eqn:transformation2}
    \vec{Y}\vec{\nabla}C(\vec{x})+\vec{X}H_{C}\vec{X}\geq0
\end{equation}

Note that as long as $|t(x_i(\varepsilon))|=|t(x_i)|$ holds, the only needed information of the transformation is $\vec{X},\vec{Y}$. As we have shown, the above equations holding is necessary, as otherwise a transformation could improve the tile. Another possibility is, that the transformation itself is not possible as an involved parameter is on a boundary and cannot be moved further. We thus can use these equations not holding for specific transformations to follow that these parameters have to be on a boundary and thus rule out specific tile structures.
\end{observation}

\begin{lemma}
    \label{lem:outer_step}
    Let $t$ be a tile with the leftmost sections
    being up to one step followed by up to one
    slide.
    Then these sections can be replaced by up to one step $s=(q_1,q_2)$ and up to one slide $h=(q_2,q_3)$ such that
    \begin{enumerate}
        \item If $s$ exists, then $x_1 x_2\ge 1/\sqrt{2}$
        \item If $h$ exists, then $x_1 x_2\le 1/\sqrt{2}$
    \end{enumerate}
    giving us a tile $\tilde{t}\preceq t$.
\end{lemma}

\begin{proof}
    We assume that both leftmost sections, i.e., a step $s=(q_1,q_2)$, followed by a slide $h=(q_2,q_3)$ exist. The other cases can be treated similarly.

    We analyze how $|t|$ and $|C_t|$ change as we move
    $q_1$ and $q_2$ along $h_t$.
    To obtain $\tilde{t}\preceq t$, we must have $|\tilde{t}|=|t|$. This is satisfied if $1+(x_2-x_1)/{x_2}+\ln(x_3/x_2)$ is left constant.
    Using the substitution $x_1=-x_2(\alpha+\ln(x_2))$, this term becomes $2+\alpha+\ln(x_3)$,
    which is independent of $x_2$.

    As we also require that the crown gets smaller, we analyze the contribution $C$ of both sections to the crown of $t$. For $D=1+\alpha+\ln(x_3)+\frac14(x_3^2-x_3^{-2})$:
    \begin{align*}
        C&=|T(q_1,q_2)|+H(q_2,q_3)\\
        &=\frac12\left(\frac1{x_1}-\frac1{x_2}+{x_2}-{x_1}\right)\left({x_1}+\frac1{x_2}\right)+
        \left[\ln(z)+\frac14(z^2-z^{-2})\right]_{x_2}^{x_3}\\
        &=D-\frac14\left(\frac{1+2(\alpha+\ln(x_2))^{-1}}{x_2^2}
        +(1+2\alpha(1+\alpha)+2\ln(x_2)(1+2\alpha+\ln(x_2)))x_2^2\right)\\
        \frac{\partial C}{\partial {x_2}}&=\frac{(1+\alpha+\ln(x_2))^2 \left((\alpha+\ln(x_2))^{-2}-2x_2^4\right)}{2x_2^3}\\
        &=\frac{(x_2-x_1)^2 (1-2x_1^2 x_2^2)}{2 x_1^2 x_2^3}
    \end{align*}
    (Note that the area is only dependent on $\alpha$ and $x_3$ and not $x_2$, so differentiation w.r.t.\ $x_2$ is justified.)
    The last term is positive iff $x_1 x_2<1/\sqrt{2}$. In this case we can decrease $x_2$ until either $x_1 x_2=1/\sqrt{2}$ or the step was completely removed.
    Similarly, if $x_1 x_2>1/\sqrt{2}$, we can increase $x_2$ until either $x_1 x_2=1/\sqrt{2}$ or the slide was completely removed. In both cases, the statement easily follows.
\end{proof}

\begin{lemma}
    \label{lem:shorter_side}
    Let $t$ be a tile with a step $(q_1,q_2)$ and a slide $(q_2,q_3)$.
    If $x_1\ge 1/x_3$, then the two sections can be replaced
    by a slide $(q_1,q_4)$ and a step $(q_4,q_2)$,
    resulting in a tile $\tilde{t}\preceq t$.
\end{lemma}

\begin{proof}
    To obtain $\tilde{t}\preceq t$, the tile has to keep its area
    upon replacing the sections, so
    $(x_2-x_1)/x_2+\ln{(x_3/x_2)}=\ln{(x_4/x_1)}+(x_3-x_4)/x_3$.
    This is fulfilled if we set $x_4=\frac{x_1 x_3}{x_2}$.

    Let $C$ be the contribution of both sections to $C_t$ and $\tilde{C}$ be the contribution of both sections to $C_{\tilde{t}}$. We are interested in the difference $\Delta$ between both:
    \begin{align*}
        C&=|T(q_1,q_2)|+H(q_2,q_3)\\
        &=\frac12\left(\frac1{x_1}-\frac1{x_2}+x_2-x_1\right)\left(x_1+\frac{1}{x_2}\right)+\left[\ln(z)+\frac14(z^2-z^{-2})\right]_{x_2}^{x_3}\\
        \tilde{C}&=H(q_1,q_4)+T(q_4,q_3)\\
        &=\left[\ln(z)+\frac14(z^2-z^{-2})\right]_{x_1}^{x_4}+\frac12\left(\frac1{x_4}-\frac1{x_3}+x_3-x_4\right)\left(x_4+\frac{1}{x_3}\right)\\
        &=\left[\ln(z)+\frac14(z^2-z^{-2})\right]_{x_1}^{\frac{x_1 x_3}{x_2}}+\frac12\left(\frac{x_2}{x_1 x_3}-\frac1{x_3}+x_3-\frac{x_1 x_3}{x_2}\right)\left(\frac{x_1 x_3}{x_2}+\frac{1}{x_3}\right)\\
        \Delta&=\tilde{C}-C\\
        &=\frac{(x_2-x_1)^2 (x_3^2-x_2^2) (1-x_1^2 x_3^2)}{4x_1^2 x_2^2 x_3^2}\\
        &\le \frac{(x_2-x_1)^2 (x_3^2-x_2^2) (1-\frac{1}{x_3^2} x_3^2)}{4x_1^2 x_2^2 x_3^2}=0
    \end{align*}
    This immediately gives us a new tile $\tilde{t}\preceq t$.
\end{proof}

\begin{lemma}
    \label{lem:step_merging}
    Let $t$ be a tile and $(q_1,q_2)$,$(q_2,q_3)$ be steps with $x_3\leq1$. Then the two steps can be replaced with a step $(q_1,q_3)$ and a slide $(q_3,q_4)$ resulting in a tile $\tilde{t}\preceq t$.
\end{lemma}

\begin{figure}
    \centering
    \begin{minipage}[c]{0.4\linewidth}
    \includegraphics[width=\linewidth]{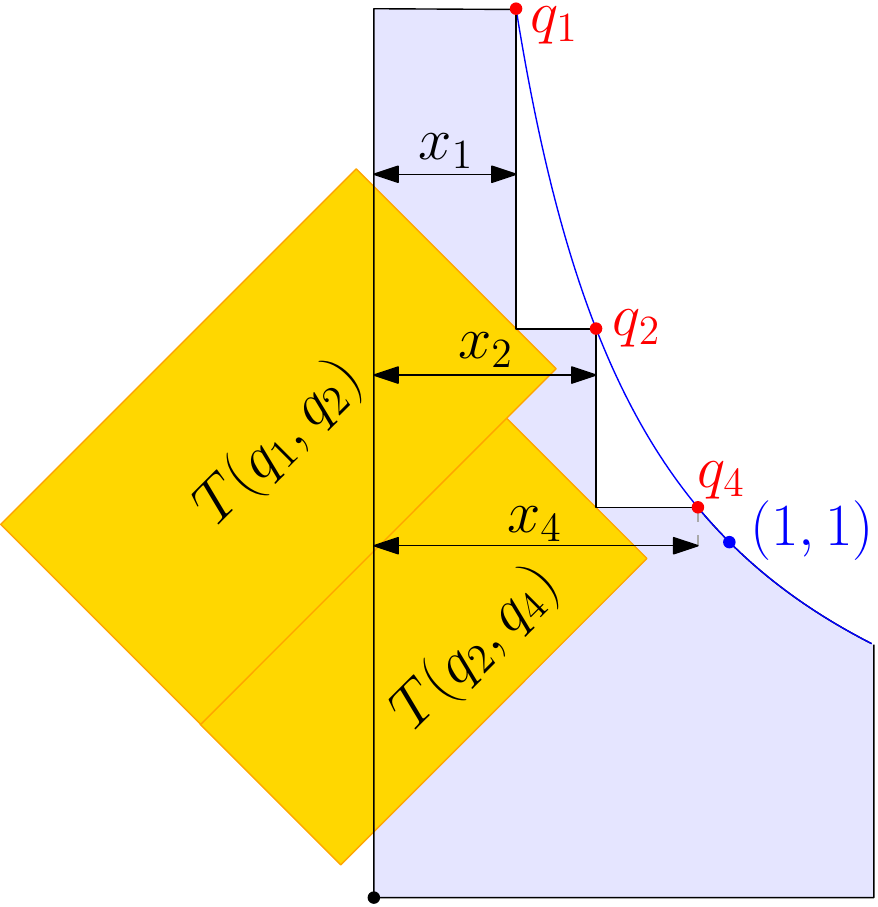}
    \end{minipage}
    \begin{minipage}[c]{0.15\linewidth}
\centering
    \cref{lem:step_merging}\par
    \includegraphics[width=\linewidth]{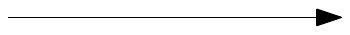}
    \end{minipage}
    \begin{minipage}[c]{0.4\linewidth}
    \includegraphics[width=\linewidth]{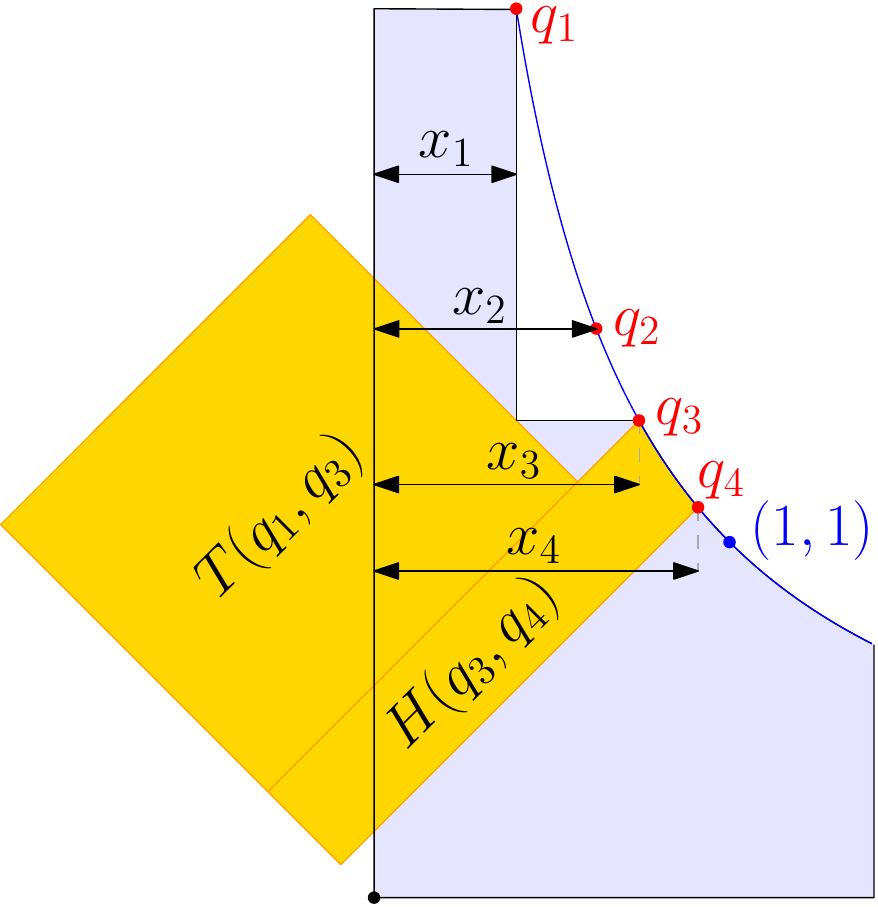}
    \end{minipage}
    \caption {
        Merging two steps in \Cref{lem:step_merging}.
    }
    \label{fig:step_merging}
\end{figure}

\begin{proof}
    See \Cref{fig:step_merging} for a depiction.
    We will look at a generalized tile $t'$ with two steps $(q_1,q_2),(q_2,q_3)$ and a slide $(q_3,q_4)$. $t$ can be understood as having $q_3=q_4$,
    such that the slide vanishes, and $\tilde{t}$ corresponds to $q_2=q_3$.
    We will move $q_3$ to the left on the hyperbola
    and let $q_2$ move such that the transformation is
    area-preserving.
    $t'$ is then obtained when both points coincide.
    Therefore we show $\partial |C_{t'}|/\partial x_3\ge 0$ while letting $x_2:=f(x_3)$ for area preservation.
    We first look at a part $u$ of the tile area which
    has to stay constant. This then also implies a formula for the derivative of $x_2=f(x_3)$ w.r.t.\ $x_3$:
    \begin{align*}
        u&:=1+(x_2-x_1)/x_2+(x_3-x_2)/x_3+\ln(x_4/x_3)\\
        &=1+(f(x_3)-x_1)/f(x_3)+(x_3-f(x_3))/x_3+\ln(x_4/x_3)\\
        \frac{\partial f(x_3)}{\partial x_3}&=\frac{f(x_3)^2 (x_3-f(x_3))}{x_3 (x_1 x_3-f(x_3)^2)}\\
        &=\frac{x_2^2 (x_3-x_2)}{x_3 (x_1 x_3-x_2^2)}
    \end{align*}
    Solving the above area equality for $x_2$ we obtain the two solutions
    \begin{align*}
        x_2^{\pm}=\frac12(y\pm\sqrt{y^2-4 x_1 x_3})
    \end{align*}
    where $y=(3-u+\ln(x_4/x_3))x_3$. Note that at least one of the solutions must be real,
    so $y^2>4 x_1 x_3$ has to hold.
    We will show that $x_2=x_2^{+}$ always yields a smaller crown area.
    To see this, look at the difference $\Delta$ of the contribution $C$ of the sections to the crown area:
    \begin{align*}
        C=&|T(q_1,q_2)|+|T(q_2,q_3)|+H(q_3,q_4)\\
        =&
        \frac12\left(\frac{1}{x_1}-\frac{1}{x_2}+x_2-x_1\right)\left(x_1+\frac{1}{x_2}\right)+
        \frac12\left(\frac{1}{x_2}-\frac{1}{x_3}+x_3-x_2\right)\left(x_2+\frac{1}{x_3}\right)+\\
        &\left[\ln z+\frac14(z^2+z^{-2})\right]_{x_3}^{x_4}\\
        \Delta=&\left.C\right|_{x_2=x_2^{+}}-\left.C\right|_{x_2=x_2^{-}}\\
        =&\frac{(1-x_1^2 x_3^2)\sqrt{y^2-4 x_1 x_3} (y-x_1-x_3)}{2 x_1^2 x_3^2}\\
        =&\frac{(1-x_1^2 x_3^2)\sqrt{y^2-4 x_1 x_3} (x_2-x_1)(x_3-x_2)/(-x_2)}{2 x_1^2 x_3^2}\\
        \le& 0
    \end{align*}
    where the last inequality is true due to $0<x_1<x_2<x_3\le 1$.

    Using $\partial f(x_3)/\partial x_3$ from above, we get for $C:=\left.C\right|_{x_2=f(x_3)}$:
    \begin{align*}
        \frac{\partial C}{\partial x_3}&=\frac{(x_3-x_2)X}{2x_1 x_2 x_3^3 (x_2^2-x_1 x_3)}
    \end{align*}
    where $X=x_2 x_3^2+x_1^2 x_3(1-x_2^3 x_3+x_2 x_3^3)+x_1(x_2 x_3-x_2^2+2(x_2^4-1)x_3^2-2x_2^3 x_3^3)$.
    The term $(x_3-x_2)/(2x_1 x_2 x_3^3 (x_2^2-x_1 x_3))$ is positive since $x_2=(y+\sqrt{y^2-4 x_1 x_3})/2\ge y/2\ge \sqrt{x_1 x_3}$, so it only remains to show $X\ge 0$.
    We apply the substitution $x_1=\alpha \beta^2 x_3$ and $x_2=\beta x_3$. The constraints $0<x_1<x_2<x_3\le 1,x_2\ge \sqrt{x_1 x_3}$ then rearrange into $0<\alpha,\beta,x_3\le 1$ and we get for $X$:
    \begin{align*}
        X=& 1-\alpha \beta (2-\beta+(1-\alpha)\beta^2)-\alpha(1-\beta)(2-\alpha-\alpha \beta)\beta^4 x_3^4\\
        \ge&  1-\alpha\beta (2-\beta+(1-\alpha)\beta^2)-\alpha(1-\beta)(2-\alpha-\alpha \beta)\\
        =& (1-\alpha)(1-\alpha(1-(1-\beta)\beta^2))\\
        \ge& (1-\alpha)^2\\
        \ge& 0
    \end{align*}
    which proves the statement.
\end{proof}

\begin{lemma}
    \label{lem:slope_lemma}
    Let $t$ be a tile. Then either there exists a tile $\tilde{t}\preceq t$ with $\ammaG_{\tilde{t}}\subset h_{\tilde{t}}$, or $t$ contains a double step $(q_1,q_2)$ with $x_1\le 1\le x_2$ or a corner $(q_1,q_2)$ with $x_1< 1$ (if $q_2\notin h_t$, $x_2>1$ if $q_1\notin h_t$).
\end{lemma}

\begin{proof}
    By assumption we only have at most one point $q\in\ammaG_t$ which lies not on the
    hyperbola. If there is no such point, then we are done, as such we assume this point exists.

    First consider the case where $q$ is part of a corner. (w.l.o.g.\ we assume it to be of the form $(q,q_1)$, where $q\notin h_t$. We then have $x_1<1$). The crown associated to $q$ is then $C_q=\frac12(x+\frac1{x_1})(y-\frac1{x_1}+x_1-x)$.
    We will move $q$ to the top left or bottom right such that the tile area does not change. This transformation is $x\rightarrow x+\frac{x (x_1 y-1)}{x_1(y + \varepsilon)-1}, y\rightarrow y+\varepsilon$, which as can be easily checked leaves the tile area invariant. Using \Cref{obs:transformation} this leads to \Cref{eqn:transformation2}: $x_1^2-3 x x_1 -1\geq 0$, which is easily checked to be false since $0<x<x_1<1$ holds.

    As such the minimal crown must be on the boundary of the transformation, i.e., where $q$ is on the hyperbola.\\
    Now assume that $q$ is part of the double step $(q_1,q,q_2)$. We denote $h_1=\frac{1}{x_1}-y,w_1=x-x_1,h_2=y-\frac1{x_2},w_2=x_2-x$.
    W.l.o.g.\ assume that $x_2<1$, we then denote by $q_0$ and $q_{1}$ the two leftmost points. As $q_0\in h_t$, we get that the crown associated with these two points is $C=\frac12(x_1-x_0+y_0-y_1)(x_0+y_1)$. It is easy to check that moving $q_0$ down decreases the crown as well as $|t|$.\christoph{reducing both doesnt prove the statement.} As such, we are only interested in cases where $h_1>h_2$ and $w_2>w_1$, as otherwise it is easy to see that we can move $q$ up or right, respectively. This decreases $|C_t|$ while increasing $|t|$. Together with moving $q_0$ down, we can thus reduce $C_t$ further while leaving $|t|$ invariant until either $q_0$ degenerates, or $q$ reaches the hyperbola. At the end this yields a corner, which we have already dealt with.

    For the remaining cases we move $q$ such that the tile area remains the same, so $w_1 h_2=(w_1+\varepsilon)(h_2+\delta)$
    (with one of $\varepsilon$,$\delta$ being negative), or equally $\delta=w_1 h_2/(w_1+\varepsilon)-h_2$ We thus obtain the transformation $h_1\rightarrow h_1-\delta$, $w_1\rightarrow w_1+\varepsilon$, $h_2\rightarrow h_2+\delta$, $w_2\rightarrow w_2-\varepsilon$.

    The crown is then given by:
    \begin{align*}
        C(\varepsilon)&=\frac12(h_1+w_1)(x_1+1/x_1-h_1)+\frac12 (h_2+w_2)(x_1+w_1+1/x_1-h_1-h_2)
    \end{align*}

    For the derivatives w.r.t.\ $\varepsilon$ at $\varepsilon=0$, we get
    \begin{align*}
        \left.\frac{\partial C(\varepsilon)}{\partial \varepsilon}\right|_{\varepsilon=0}&=\frac12\left(\frac{h_2(h_2-h_1)}{w_1}-w_1+w_2\right)\\
        \left.\frac{\partial^2 C(\varepsilon)}{\partial \varepsilon^2}\right|_{\varepsilon=0}&=\frac{(h_1-2h_2)h_2}{w_1^2}-1
    \end{align*}

    Thus, we can now conclude $h_2<w_2$, as otherwise we get for the first derivative
    $\frac12\left(\frac{h_2(h_2-h_1)}{w_1}-w_1+w_2\right)\le \frac12\left(w_2-w_1-\frac{h_2^2}{w_1}\right)<0$ since
    $w_2\le h_2$ and $w_1<w_2\le h_2$. As we thus find a direction in which our transformation lowers the crown area.

    Additionally, if the second derivative is negative, $C(\varepsilon)$ has no minimum and thus the minimum has to be on the boundary.
    So it remains to handle the case where $(h_1-2h_2)h_2\ge w_1^2$, $h_1>h_2$, $w_2>w_1$ and $w_2>h_2$.
    We move $q$ to its new location $q'=(x_2-(h_2-\gamma), y_2+(h_2-\gamma))$. Since  $x_2<1$, $q'$ remains below the hyperbola. $\gamma$ can be chosen such that $A=B$: For $\gamma=0$, we have $A=0<(w_2-h_2)h_2=B$ and for $\gamma=h_2$ we have $A=w_1 h_2>0=B$. Since $A$ and $B$ change continuously
    while we change $\gamma$ in $[0,h_2]$, there must be some $\gamma$ where $A=B$ and therefore $\gamma w_1=(h_2-\gamma)(w_2-(h_2-\gamma))$. Note that $w'_2>w'_1$ reduces to $\gamma<h_2-\frac{w_1}2-\frac{w_2}2$, which can be easily checked to be false. As such the resulting tile can be dealt with by the above method, and it remains to show that crown area is reduced:
    \begin{align*}
        \Delta&=\frac12(h_1'+w_1')(x_1+1/x_1-h_1')+\frac12 (h_2'+w_2')(x_1+w_1'+1/x1-h_1'-h_2')
\\&-\frac12(h_1+w_1)(x_1+1/x_1-h_1)-\frac12 (h_2+w_2)(x_1+w_1+1/x_1-h_1-h_2)\\
&=\frac12 (h_1 - w_1 + 3 h_2) (h_2 - w_2) + (-h_1 + w_1 - 3 h_2 + w_2) \gamma + \gamma^2
\end{align*}
    It is easily seen that $\partial \Delta/\partial h_1<0$ holds, as $\gamma$ is independent of $h_1$. As we want to show $\Delta<0$ we thus can assume $h_1$ to be minimal, i.e., $h_1=w_1^2/h_2+2h_2$. We thus obtain:
    \begin{align*}
        \Delta\leq&\frac{(w_1^2 - w_1 h_2 + 5 h_2^2) (h_2 - w_2)}{2 h_2} + w_1 \gamma - \frac{
 w_1^2 \gamma}{h_2} + (-5 h_2 + w_2) \gamma + \gamma^2
    \end{align*}
    It is more involved but still straightforward to check that the derivative of this term with respect to $w_2$ is also negative. (Note that $\gamma$ is dependent on $w_2$). Thus choosing $w_2$ minimal, i.e., $w_2=h_2$ we obtain:
    \begin{align*}
        \Delta\leq& \gamma (w_1 - \frac{w_1^2}{h_2} - 4 h_2 + \gamma)\leq0
    \end{align*}
    where the last inequality follows from $w_1\le w_2=h_2$ and $0<\gamma<h_2$. With this we can always find $\tilde{t}$ with the required properties.
\end{proof}

\begin{lemma}
    \label{lem:enforced_high_density}
    Consider a tile $t$ with $\rho_t\leq 1/2$. Then there exists a tile  $\tilde{t}\preceq t$, containing at most one step. Furthermore, if $\ammaG_t\subseteq h_t$ then $\ammaG_{\tilde{t}}\subseteq h_{\tilde{t}}$.
\end{lemma}

\begin{proof}
    If $t$ only contains one step we are already done. Otherwise, we denote by $s=(q_1,q_2)$ and $s'=(q_3,q_4)$ the leftmost and rightmost step respectively.
    Assume that no transformations are available from \Cref{lem:step_merging}, \Cref{lem:shorter_side} and \Cref{lem:slope_lemma}. We show that $\rho_t>1/2$ with a contradiction.

    Assume that $q_2\neq q_3$, then at least one section has to exist between $s$ and $s'$. If we assume $1<x_1<x_2<x_3<x_4$ (or its mirror), \Cref{lem:step_merging} and \Cref{lem:shorter_side} rule out a step or slide, and \Cref{lem:slope_lemma} exclude a corner/double-step. As such we can assume that $x_1<1<x_4$. We then differentiate between the cases $x_1<1<x_2$ (or equivalently its mirror) and $x_2\leq1\leq x_3$.
    In the first case, there can only be one slide between $s$ and $s'$. In both cases attaching a slide before $s$ or after $s'$ is impossible: Either we get a slide-step-slide sequence, which is impossible by \Cref{lem:shorter_side}, or we get a step-slide sequence, which is impossible by \Cref{lem:shorter_side}.
    As such $s'$ and $s$ are both boundary steps. Since \Cref{lem:outer_step} was inapplicable, we get that $x_1 x_2\geq 1/\sqrt{2}$ and $y_3 y_4\geq 1/\sqrt{2}$. From $x_1<x_2$ and $y_4<y_3$ we then get $1/\sqrt{2}<x_2^2$ and $1/\sqrt{2}<y_3^2$. With this we bound $|t|$ from above:
    \begin{align*}
        |t|\le& 1+(x_2-x_1)/x_2+\ln(x_3/x_2)+(x_4-x_3)/x_4\\
        =&3-x_1/x_2-\ln(y_3 x_2)-y_4/y_3\\
        \le&3-(x_2^{-2}/\sqrt{2}+\ln x_2)-(y_3^{-2}/\sqrt{2}+\ln y_3)
    \end{align*}
    The term $z^{-2}/\sqrt{2}+\ln z$ is minimized at $z=\sqrt[4]{2}$ with value $(2+\ln 2)/4$. We conclude $\rho_t=1/|t|\ge 1/(3-2(2+\ln 2)/4)\approx 0.6>1/2$.

    Instead suppose $q_2=q_3$. We get again that $x_1<1<x_4$ by \Cref{lem:step_merging}.
    This means that \Cref{lem:slope_lemma} does not allow the existence of a corner/double-step. So $s, s'$ are the outermost steps, and only slides can exist beside $s$ and $s'$. If no slide exists, we end up in the case above. By symmetry we can w.l.o.g.\ choose $x_1<1\leq x_2$, which means that \Cref{lem:shorter_side} does not permit a slide after $s'$. Therefore a slide $h=(q_0,q_1)$ exists. By \Cref{lem:shorter_side} we know that $x_0>1/x_2$. Since $s'$ is a boundary step, \Cref{lem:outer_step} gives us $1/(x_2 x_4)\geq 1/\sqrt{2}$. The total area of $|t|$ is then:
    \begin{align*}
        |t|=& 1+\ln(x_1/x_0)+(x_2-x_1)/x_2+(x_4-x_2)/x_4\\
        =&3-x_1/x_2-x_2/x_4+\ln(x_1/x_0)\\
        \le &3-x_1/x_2-x^2_2/\sqrt{2}+\ln(x_2 x_1)\\
        \le &3-1/x_2-x^2_2/\sqrt{2}+\ln(x_2)\leq 1.4
    \end{align*}
    where the second to last inequality follows from
    the derivative w.r.t.\ $x_1$ being $1/x_1-1/x_2>0$, so $x_1$ was maximized. We conclude $\rho_t=1/|t|>1/2$.

    Therefore we conclude that we can use transformations of \Cref{lem:step_merging}, \Cref{lem:shorter_side} and \Cref{lem:slope_lemma} to reduce the step count to $\leq 1$. As none of these transformations moves points in $\ammaG_t$ away from the hyperbola, this gives the second property. This gives us the required tile $\tilde{t}\preceq t$.
\end{proof}

\begin{lemma}
    \label{lem:inner_nonhyp_point_or_hyperbola}
    Let $t$ be a tile with a double step $(q_1,q_2,q_3)$ and a slide $(q_3,q_4)$.
    We can replace both sections by a double step $(q_1,q_4)$ or a sequence of steps and slides between $q_1$ and $q_4$, obtaining a tile $t'\preceq t$.
\end{lemma}

\begin{proof}
    We have that $|t_R(\vec{q})|=1 + (x_2 - x_1)y_2 + (x_3 - x_2)/x_3 + \log(x_4/x_3)$ and $C_{t_R}(\vec{q})=T(q_1,q_2)+T(q_2,q_3)+H(q_3,q_4)=\frac12 (\frac{1}{x_1} - x_1 + x_2 - y_2) (x_1 + y_2)+\frac12 (x_2 + \frac{1}{x_3}) (-x_2 - \frac{1}{x_3} + x_3 + y_2)+\frac14(-\frac{1}{x_4^2} + x_4^2 + \frac{1}{x_3^2} - x_3^2+ 4 \log(\frac{x_4}{x_3}))$. Where $t_R$ is the part of the tile consisting of the double step/slide.
    We will transform the tile by moving $q_2$ up/down while moving $q_3$ on the hyperbola to leave the tile area invariant.
    We define this transformation by $(y_2\rightarrow y_2+\varepsilon,x_3\rightarrow-x_2/W(-\frac{e^{-x_2/x_3 + x1 \varepsilon - x_2 \varepsilon} x_2}{x_3}))$, where $W$ denotes the main branch of the product log function. This transformation leaves the rest of the tile invariant, and it is thus straightforward to check, that $|t(\vec{q}(\varepsilon))|=|t(\vec{q})|$ holds, which as outlined in \Cref{obs:transformation} leads to \Cref{eqn:transformation2}
    \begin{align*}
        \Delta:&=4 x_2 x_3^2 - 2 x_3^3 - 2 x_2^4 x_3^3 + x_2^3 (1 + 4 x_3^4) -
 x_2^2 x_3 (4 + 2 x_3^4 - x_3 y_2) \\&-
 2 x_1 x_3^2 (-1 + x_2 (-2 (x_2 - x_3)^2 x_3 + y_2)) -
 x_1^2 (2 x_2^2 x_3^3 - x_2 (1 + 4 x_3^4) \\&+ x_3 (2 + 2 x_3^4 - x_3 y_2))\geq0
    \end{align*}
    We will show that this equation does not hold.
    It is easy to check that $\frac{\partial \Delta}{\partial y_2 }=(x_1 - x_2)^2 x_3^2>0$ holds, meaning that $\Delta$ is maximal for big $y_2$. Note that $x_2 y_2<1$ has to hold, since $q_2$ is not on the hyperbola, thus, inserting $y_2=1/x_2$ we get that $\Delta'=x_1^2 + x_2^2 - 2 x_2 x_3 - 2 (x_1 - x_2)^2 x_2 x_3^3\geq0$ has to hold. As $x_1<x_2<x_3$ holds, $\frac{\partial \Delta'}{\partial x_1 }=2 x_1 + 4 x_2 (-x_1 + x_2) x_3^3>0$ holds as well. Leading to $x_1=x_2$ as the maximal possible value for $\Delta'$, we obtain $\Delta'|_{x_1=x_2}=2 x_2 (x_2 - x_3)<0$, contradicting \Cref{eqn:transformation2}. As the equation is a necessary condition for the transformation not being able to lower $C_t$, the transformation is thus always possible, unless either the slide is exhausted or $q_2$ reaches the hyperbola/degenerates.
\end{proof}

\begin{lemma}
    \label{lem:hyp_merge_with_boundary_nh_step}
    Let $t$ be a tile with a slide $(q_1,q_2)$ and a corner $(q_2,q_3)$.
    We can replace both sections by
    a corner $(q_1,q_2')$ or a sequence of steps and slides between $q_1$ and $q_2'$, obtaining a tile $t'\preceq t$.
\end{lemma}

\begin{proof}
    We have $|t_R(\vec{q})|=1 + \log(x_2/x_1) + (x_3 - x_2) y_3$ and
$C_{t_R}(\vec{q}) =H(q_1,q_2)+T(q_2,q_3)= \frac14(1/x_1^2 - x_1^2 - 1/x_2^2 + x_2^2 + 4 \log(x_2/x_1))+\frac12(1/x_2 - x_2 + x_3 - y_3) (x_2 + y_3) $. Where $t_R$ is the part of the tile consisting of the corner/slide.
    We define the transformation $x_2\rightarrow x_2+k\varepsilon$,$x_3\rightarrow x_3+\beta$,$y_3\rightarrow y_3+\varepsilon$. This transformation leaves the tile besides $t_R$ invariant, and as such requiring preservation of tile area, leads to $\beta= \varepsilon + k \varepsilon^2 + \log(x_2) - \log(x_2 (x_2 \varepsilon - x_3 \varepsilon + k y_3+ k \varepsilon))/(y_3 + \varepsilon)$. Thus $|t(\vec{q}(\varepsilon))|=|t(\vec{q})|$ holds, and we can follow \Cref{obs:transformation}, and analyze the resulting equations. Note that the defined transformation is dependent on $k$, and that the equations have to be satisfied for every $k$, as otherwise there is a $k$ defining a valid transformation which reduces $C_t$.

    For $k=0$, this results for \Cref{eqn:transformation1} in $x_2 (x_2 + (x_2 (x_3 - x_2))/y_3 + 2 y_3) = 1$, which reduces to $x_3=(x_2^3 + y_3 - x_2^2 y_3 - 2 x_2 y_3^2)/x_2^2$. Inserting this into \Cref{eqn:transformation2}, again for $k=0$, we obtain $y_3 < (1 - x_2^2)/(3 x_2)$. Returning to a general $k$ for \Cref{eqn:transformation2}, we obtain a quadratic polynomial in $k$. The coefficient of $k^2$ then has to be positive, as otherwise we can chose $k$ big enough so that \Cref{eqn:transformation2} is no longer satisfied. This results in $-x_2^3 + (x_2^4 - x_2^2 - 3) y_3 + 2 x_2 y_3^2>0$ having to hold. It is easy to check that the left-hand side has positive curvature for $y_3$, so only the boundary values ($0<y_3 < (1 - x_2^2)/(3 x_2))$ are of interest. For $y_3=0$ the equation does not hold, and inserting $y_3=(1 - x_2^2)/(3 x_2)$ yields $-(7 - 2 x_2^2 + x_2^4 + 3 x_2^6)/(9 x_2)>0$, which has a single maximum for $x_2^2\approx0.767$ at $\approx-0.94$, so the equation does not hold in general.
    As the equation is a necessary condition for the transformation not being able to lower $C_t$, the transformation is thus always possible for some $k$, until either the slide is exhausted or $q_3$ reaches the hyperbola/degenerates.
\end{proof}
\begin{lemma}\label{lem:hilfslemma}
Let $t$ be a tile with $\rho_t\leq 1/2$, then we can find a tile $\tilde{t}\preceq t$ with one of the following properties:
\begin{itemize}
    \item $\ammaG_{\tilde{t}}\subseteq h_{\tilde{t}}$
    \item $\tilde{t}$ consists of only a step and a double step
    \item $\tilde{t}$ consists of only a double step
    \item $\tilde{t}$ consists of only a step and a corner
\end{itemize}
\end{lemma}
\begin{proof}
By \Cref{lem:enforced_high_density} we first assume one step exists. If $|\ammaG_t\setminus h_t|=0$, we are done, so we assume $|\ammaG_t\setminus h_t|=1$. $t$ then has to contain a double step or corner, which by \Cref{lem:slope_lemma} is located around $1$. We can assume that any existing slide is adjacent to the double-step/corner. This is true as otherwise the sections have to be separated by the step, meaning we can use \Cref{lem:shorter_side} to exchange the step and the slide to make them adjacent, obtaining a tile $\tilde{t}\preceq t$. By this adjacency we can use \Cref{lem:inner_nonhyp_point_or_hyperbola} or \Cref{lem:hyp_merge_with_boundary_nh_step} to either obtain $|\ammaG_t\setminus h_t|=0$, or the nonexistence of slides. As such our tile can only consist of a step and a double-step/corner. The possibility of $t$ just consisting of a corner can be excluded, as its density would be larger than $1/2$. With this we obtain the result.
\end{proof}

\begin{lemma}
    \label{lem:step_plus_hyperbola_section}
    Let $t$ be a tile with $\rho_t\le 1/2$
    with $\ammaG_t\subseteq h_t$.
    Then $|C_t|/|t|\ge \LBii(\rho_t)$.
\end{lemma}

\begin{proof}
    By \Cref{lem:enforced_high_density} we can assume
    that $t$ contains at most one step.
    Since $\ammaG_t\subseteq h_t$, $t$ can only consist of steps and slides.
    Using \Cref{lem:shorter_side}, we can ensure that $t$ has
    at most one slide. (Note that a slide must exist, as otherwise $\rho_t>1/2$ holds.)
    It is enough to show $\Delta:=|t| \LBii(\rho_t)-|C_t|<0$, since the statement follows by rearranging.

    First assume that $t$'s only section is a slide $(q_1,q_2)$. Then $|t|=1+\ln(x_2/x_1)$,
    and we get:
    \begin{align*}
        \Delta&=|t| \LBii(\rho_t)-H(q_1,q_2)\\
        &=|t|-1+\sinh\left(|t|-1\right)-\left[\ln z+(z^2-z^{-2})/4\right]_{x_1}^{x_2}\\
        &=\left(x_2^{-2}-x_2^2+x_1^2-x_1^{-2}\right)/4+\sinh\left(|t|-1\right)+|t|-\left(1+\ln\left(x_2/x_1\right)\right)\\
        &=\left(x_2^{-2}-x_2^2+x_1^2-x_1^{-2}\right)/4+\sinh\left(\ln\left(x_2/x_1\right)\right)\\
        &=2\sinh\left(\left(\ln\left(x_1\right)+\ln\left(x_2\right)\right)/2\right)^2\sinh\left(\ln\left(x_1/x_2\right)\right)<0
    \end{align*}
    where the last inequality directly follows from $x_1<x_2$.

    Now assume that $t$ consists of a step $(q_1,q_2)$ and a slide $(q_2,q_3)$ (w.l.o.g.\ ordered in this way).
    In such a case we have $|t|=1+(x_2-x_1)/{x_2}+\ln(x_3/x_2)$, or rearranged, $x_3=x_2 e^{x_1/x_2+|t|-2}$.
    Again we calculate
    \begin{align*}
        \Delta&=|t| \LBii(\rho_t)-(|T(q_1,q_2)|+H(q_2,q_3))\\
        &=|t|-1+\sinh(|t|-1)-\frac12\left(\frac{1}{x_1}-\frac{1}{x_2}+x_2-x_1\right)\left(x_1+\frac{1}{x_2}\right)-\left[\ln z+\frac{z^2-z^{-2}}{4}\right]_{x_2}^{x_3}
    \end{align*}
    Taking the derivative of $\Delta$ w.r.t.\ $|t|$ (after inserting $x_3=x_2 e^{x_1/x_2+|t|-2}$), we obtain
    $\partial \Delta/{\partial |t|}=-2\cosh\left(x_1/x_2+|t|-2+\ln(x_2)\right)^2<0$. This indicates that $\Delta$ is maximized for smallest $|t|$. So assume $|t|=2$ now, or equivalently, $x_3=x_2 e^{x_1/x_2}$. By \Cref{lem:outer_step} we can further assume that $x_2=1/(\sqrt{2}x_1)$. Using the substitution $x_1=2^{-3/4}\sqrt{u}$ we get
    \begin{align*}
        \Delta&=\frac{1}{2\sqrt{2}}\left(-3-\frac{\sqrt{2}}{e}+\sqrt{2}e+\frac{1-e^u}{u}+u+\frac{u}{2e^u}\right)\\
        \frac{\partial \Delta}{\partial u}&=\frac{(1-u)(u^2+2e^{2u}-2e^u(1+u))}{4\sqrt{2}u^2 e^u}
    \end{align*}
    The derivative above has only one zero, namely $u=1$:
    From $0<x_1^2\le x_1 x_2=1/\sqrt{2}$, we can deduce $u\in\intoc{0,2}$ by the substitution.
    For the right factor of the derivative's numerator we then get
    $u^2+2e^{2u}-2e^u(1+u)>2e^u(e^u-(1+u))>0$ from the Taylor series of $e^u$.
    Hence checking $\Delta$ at $u$'s boundaries and $u=1$ is sufficient, where we get $\lim_{u\rightarrow 0} \Delta\approx -0.24<0$ (apply L'Hospital's rule on $(1-e^u)/u$), $\left.\Delta\right|_{u=1}\approx -0.07<0$ and $\left.\Delta\right|_{u=2}\approx -0.26<0$.
\end{proof}

The following \Cref{lem:area_two_or_all_points_on_ht} allows us to restrict ourselves to tiles $t$ with $\rho_t=1/2$ or where all points in $\ammaG_t$ are on
$t$'s hyperbola.

\begin{lemma}
    \label{lem:area_two_or_all_points_on_ht}
    Let $t$ be a tile with $\rho_t\le 1/2$. Then there
    exists a tile $\tilde{t}$ with $\ammaG_{\tilde{t}}\subseteq h_{\tilde{t}}$ or
    $\rho_{\tilde{t}}=1/2$ such that, if
    $|C_{\tilde{t}}|/|\tilde{t}|\ge \LBii(\rho_{\tilde{t}})$ then also
    $|C_t|/|t|\ge \LBii(\rho_t)$.
\end{lemma}

\begin{proof}
    If $\ammaG_t\subseteq h_t$, then the statement is trivial using $\tilde{t}=t$.
    Otherwise let $\set{q_i}=\ammaG_t\setminus h_t$.
    W.l.o.g.\ assume that there is a point in $q_{i-1}\in\ammaG_t$ with $x_{i-1}<x_i$ (otherwise look at the symmetric case; another point has to exist since $h_t\cap \ammaG_t \ne \emptyset$).
    We construct a new tile $\tilde{t}$ with $\rho_{\tilde{t}}\le 1/2$ by moving $q$ up by $\varepsilon$. This changes the area, i.e., $|\tilde{t}|=|t|+\varepsilon w$ for $w=x_i-x_{i-1}$.
    It is then sufficient to show $|C_t|-\LBii(\rho_t)|t|>|C_{\tilde{t}}|-\LBii(\rho_{\tilde{t}})|\tilde{t}|$.

    To see this, consider the term $|C_t|-\LBii(\rho_t)|t|=|C_t|-(|t|-1+\sinh(|t|-1))=:X$ as we move $q$.
    Since $1$ or $2$ towers are affected by moving $q$ (depending on whether $q$ is part of a double step or a corner), the contribution
    of those towers to the curvature of $X$ is $-1$ each by \Cref{obs:upperstaircasemove:crown_change} using $\alpha=1$.
    Hence
    \begin{align*}
        \frac{\partial X^2}{\partial \varepsilon^2}&=
        \frac{\partial^2 |C_t|}{\partial \varepsilon^2}-\frac{\partial}{\partial \varepsilon^2} (|t|+\varepsilon w-1+\sinh(|t|+\varepsilon w-1))\\
        &=-1\cdot\LDAUOmicron{1}-w^2\sinh(|t|-1+\varepsilon w)\\
        &=-1\cdot\LDAUOmicron{1}-w^2\sinh(|\tilde{t}|)<0
    \end{align*}
    As $\frac{\partial X^2}{\partial \varepsilon^2}<0$, $\varepsilon$'s sign can be
    chosen such that $X=|C_t|-\LBii(\rho_t)|t|$ decreases (hence we obtain $\tilde{t}$ with $|C_t|-\LBii(\rho_t)|t|>|C_{\tilde{t}}|-\LBii(\rho_t)|\tilde{t}|$).

    $\varepsilon$'s magnitude can be chosen such that $q_i$ is moved up onto the
    hyperbola or $q_i$ is moved down and the tile degenerates, which can be transformed into a new tile $\tilde{t}$ by \Cref{lem:transform:degenerate}
    with $\tilde{t}\preceq t$. We assumed that $\rho_{\tilde{t}}\le 1/2$, so the transformation might stop when the density becomes exactly $1/2$ (which is supported by the lemmas statement).
\end{proof}

\begin{lemma}
    \label{lem:double_step}
    Let $t$ be a tile with $\rho_t\le 1/2$ consisting only
    of a double-step $(q_1,q_2,q_3)$.
    Then $|C_t|/|t|\ge \LBii(\rho_t)$.
\end{lemma}

\begin{proof}
    By \Cref{lem:area_two_or_all_points_on_ht} we may assume $\rho_t=\frac12$ (or $|t|=2$). We can express this in terms
    of $x_i$, $y_i$ to get $|t|=1+(x_2-x_1)y_2+(\frac1{y_3}-x_2)y_3=2$, or equivalently,
    $x_2 y_2=x_2 y_3+x_1 y_2$.
    With the substitution
    $\chi=x_2 y_2$,$\gamma=x_2/y_2$,$\alpha=y_3/y_2$,$\beta=x_1/x_2$, or rearranged,
    $x_1=\beta\sqrt{\gamma\chi}$,$x_2=\sqrt{\gamma\chi}$,$y_2=\sqrt{\chi/\gamma}$,$y_3=\alpha\sqrt{\chi/\gamma}$,
    this equation becomes $\chi(1-\alpha-\beta)=0$.
    We get for $C_t$:
    \begin{align*}
        |C_t|&=|T(q_1,q_2)|+|T(q_2,q_3)|\\
        &=\frac12\left(\frac1{x_1}-y_2+x_2-x_1\right)\left(x_1+y_2\right)+\frac12\left(y_2-y_3+\frac1{y_3}-x_2\right)\left(x_2+y_3\right)\\
        &=1+\frac12\left(\gamma\frac1\alpha+\frac{1}{\gamma}\frac{1}{\beta}+\frac1{\gamma}\chi\alpha+\gamma\chi\beta-\frac1\gamma\chi\alpha^2-\gamma\chi\beta^2-\frac1\gamma\chi-\gamma\chi\right)+\chi\left(1-\alpha-\beta\right)\\
        &=1+\frac12\left(\gamma\frac1\alpha+\frac{1}{\gamma}\frac{1}{\beta}+\frac1{\gamma}\chi\alpha+\gamma\chi\beta-\frac1\gamma\chi\alpha^2-\gamma\chi\beta^2-\frac1\gamma\chi-\gamma\chi\right)\\
        \frac{\partial |C_t|}{\partial \chi}&=\frac12\left(\frac{1}\gamma\left(\alpha\left(1-\alpha\right)-1\right)+\gamma \left(\beta\left(1-\beta\right)-1\right)\right)<0
    \end{align*}
    where the last inequality is true since $\alpha,\beta\in(0,1)$ and $\gamma>0$ holds.
    (Note that the derivative w.r.t.\ $\chi$ is justified as the area inequality is independent of $\chi\ne 0$.)
    From this we obtain a tile $t'\preceq t$ (namely the one with $\chi=1$ and same $\alpha,\beta$). Since $\chi=1$, $t'$ is consists of exactly two steps. By \Cref{lem:enforced_high_density}, we get another tile $t''\preceq t'$ with at most one step and $\ammaG_{t''}\subseteq h_{t''}$. The statement then directly follows from \Cref{lem:step_plus_hyperbola_section}.
\end{proof}

\begin{lemma}
    \label{lem:step_plus_double_step}
    Let $t$ be a tile with $\rho_t\le 1/2$ consisting only
    of a double step $(q_1,q_2,q_3)$ and a step $(q_3,q_4)$.
    Then $|C_t|/|t|\ge \LBii(\rho_t)$
\end{lemma}

\begin{proof}
    Then $|t(\vec{q})|=1+(x_2-x_1)y_2+(x_3-x_2)/x_3+(x_4-x_3)/x_4$ and
    $|C_t(\vec{q})|=T(q_1,q_2)+T(q_2,q_3)+T(q_3,q_4)=1/2\cdot (1/x_1-y_2+x_2-x_1)(x_1+y_2)+1/2\cdot (y_2-1/x_3+x_3-x_2)(x_2+1/x_3)+1/2\cdot (1/x_3-1/x_4+x_4-x_3)(x_3+1/x_4)$.

    We define the transformation $(x_2\rightarrow x_2+\varepsilon, x_4\rightarrow x_4+\delta)$.
    We then require $|t|$ to be left invariant under this transformation, which is true for $\delta=x_4(-1+x_3^2/(x_3^2+x_4(x_3 y_2 -1)\varepsilon))$.
    Following \Cref{obs:transformation} for this transformation we obtain for \Cref{eqn:transformation1}:
    \begin{align*}
    y_2=1/x_3 + (x_3^2 (x_1 - 2 x_2 + x_3) x_4)/(-x_4 + x_3 (2 + x_3 x_4^3))
    \end{align*}

    Inserting this in \Cref{eqn:transformation2} we obtain

    {\small
\begin{align*}
        \frac{-4 x_3^2 + 4 x_3 x_4 - (1 + x_3^2 (x_1 - 2 x_2 + x_3)^2) x_4^2 -
 4 x_3^3 x_4^3 + 2 x_3^2 x_4^4 + x_3^3 (x_1 - 2 x_2 + x_3)^2 x_4^5 -
 x_3^4 x_4^6}{(x_4 - x_3 (2 + x_3 x_4^3)^2}\geq0
    \end{align*}}%
    As the denominator is trivially positive, this is equivalent to:
    {\small
    \begin{align*}
        &4 x_3^2 - 4 x_3 x_4 + (1 + x_3^2 (x_1 - 2 x_2 + x_3)^2) x_4^2 + 4 x_3^3 x_4^3 -
 2 x_3^2 x_4^4 - x_3^3 (x_1 - 2 x_2 + x_3)^2 x_4^5 + x_3^4 x_4^6\leq0
    \end{align*}}%
    The curvature of the term with respect to $x_2$ is easily checked to be $-8 x_3^2 x_4^2(x_3 x_4^3-1)<0$, as we try to prove the term to be positive, we are thus only interested in the boundary values of $x_2$ ($x_1<x_2<x_3$). The resulting term turns out to be the same for $x_2=x_1$ and $x_2=x_3$, namely $4 x_3^2-4 x_3 x_4 +(1+(x_1-x_3)^2 x_3^2) x_4^2 + 4 x_3^3 x_4^3 -2 x_3^2 x_4^4 -(x_1-x_3)^2 x_3^3 x_4^5 + x_3^4 x_4^6$. This terms curvature with regards to $a$ is $-2x_3^2 x_4^2(x_3 x_4^3-1)<0$, meaning again only the boundary $0<x_1<x_3$ is relevant. For $x_1=x_3$ we obtain $(x_4-x_3(2+ x_3 x_4^3))^2>0$ and for $x_1=0$ we get $4x_3^2-4x_3x_4+x_4^2+x_3^4x_4^2+4x_3^3x_4^3-2x_3^2x_4^4-x_3^5x_4^5+x_3^4x_4^6$. The third derivative of this w.r.t.\ $x_4$ is $12x_3^2(2x_4-x_3)(5x_3^2x_4^2-2)>0$, meaning the second derivative will be minimal for minimal $x_4=x_3$, with this the second derivative is $2(1+x_3^4+5x_3^8)>0$, meaning again only the boundary terms of $x_4$ ($x_3<x_4<\infty)$ are relevant. For these we get $x_3^2+3x_3^6>0$ and $+\infty>0$ respectively. Meaning the term is always positive and thus the transformation is always possible unless either $q_2$ degenerates/hits the hyperbola, or $q_4$ degenerates. We thus can transform $t$ such that either $\Gamma_{t}\subseteq h_{t}$, or consists of just a double step. In the former case \Cref{lem:step_plus_hyperbola_section} gives the required property. In the latter case \Cref{lem:double_step}.
\end{proof}

\begin{lemma}
    \label{lem:step_plus_boundary_nh_step}
    Let $t$ be a tile with $\rho_t\le 1/2$
    consisting only of a step $(q_1,q_2)$ where $x_1 x_2\ge 1/\sqrt{2}$ and a corner
    $(q_2,q_3)$.
    Then there exists a tile $t'\preceq t$ only consisting of two steps.
\end{lemma}

\begin{proof}
    Let $x_4:=1/y_3$.
    We will first attempt to move
    $q_3$ such that $|t|=1+(x_2-x_1)/x_2+(x_3-x_2)/x_4$ is invariant, or equivalently, $x_4=(x_3-x_2)/u$ for some constant $u$. In order for $|C_t|$ to decrease, it suffices to look at $|T(q_2,q_3)|$:
    \begin{align*}
        |T(q_2,q_3)|&=\frac12\left(\frac{1}{x_2}-\frac{1}{x_4}+x_3-x_2\right)\left(x_2+\frac1{x_4}\right)\\
        &=\frac12\left(\frac{1}{x_2}-\frac{u}{x_3-x_2}+x_3-x_2\right)\left(x_2+\frac{u}{x_3-x_2}\right)\\
        \frac{\partial |T(q_2,q_3)|}{\partial x_3}&=\frac{x_2(2+x_2 x_4(1+(x_3-x_2)x_4))-x_4}{2x_2 (x_3-x_2)x_4^2}
    \end{align*}
    As we assume that $q_3$ cannot be moved such that $|C_t|$ decreases, we assume
    that the last term is $0$, or equally $(x_3-x_2)/x_4=(x_4-x_2(2+x_2 x_4))/(x_2^2 x_4^3)$.
    Since $x_1\ge\frac{1}{\sqrt{2}x_2}$ and $x_2\ge \sqrt{x_2 x_1}\ge 2^{-1/4}$, we have
    \begin{align*}
        |t|&=1+(x_2-x_1)/x_2+(x_3-x_2)/x_4\\
        &\le 1+(x_2-1/(\sqrt{2} x_2))/x_2+(x_4-x_2(2+x_2 x_4))/(x_2^2 x_4^3)\\
        &=2+x_2^{-2}(x_4^{-2}-2^{-1/2})-2x_2^{-1} x_4^{-3}-x_4^{-2}\\
        &<2+2^{1/4} x_2^{-1}(x_4^{-2}-2^{-1/2})-2x_2^{-1} x_4^{-3}\\
        &=2+x_4^{-3} x_2^{-1}(2^{1/4}x_4-2-2^{-1/4} x_4^3)\\
        &\le 2+x_4^{-3} x_2^{-1}(2\sqrt{6}/9-2)<2
    \end{align*}
    This contradicts the assumption that $\rho_t\le 1/2$, so $q_3$ must be movable until
    it lies on the hyperbola, where we get the proclaimed tile $t'\preceq t$.
\end{proof}

We are now ready to show \cref{prop:LBii}.

\begin{proof}[Proof of \cref{prop:LBii}]
By \Cref{lem:general_lower_bound} we already get the result for $\rho_t> 1/2$. For tiles with $\rho_t\leq 1/2$ we, by \Cref{lem:hilfslemma}, only have to cover the following cases: \begin{itemize}
    \item $\ammaG_{t'}\subseteq h_{t'}$. \Cref{lem:step_plus_hyperbola_section} yields the result.
    \item $t$ consists of only a double step. \Cref{lem:double_step} yields the result.
    \item $t$ consists of a step and a double step. \Cref{lem:step_plus_double_step} yields the result.
    \item $t$ consists of a step and a corner. \Cref{lem:step_plus_boundary_nh_step} yields the result.
\end{itemize}
As such the bound follows.

It remains to show the tightness.
Note that $\LBii$ exactly corresponds to the tiles
$\wcTileHD{s}$ and $\wcTileLD{s}$ shown in \Cref{fig:extreme_case_low_density}
and \Cref{fig:extreme_case_high_density}: Both tiles are symmetric tiles.
First note that for $\rho_t=1$, the tightness is trivial by choosing an arbitrary tile $t$ with exactly one upper staircase point.

Let $t=\wcTileHD{s}$ and define $\ammaG_t=\set{q_1,q_2}$ with $q_1=(v,1/v)$ and $q_2=(1/v,v)$ for proper $v\in (0,1)$.
We have $|t|=2-v^2$, or rearranged $v=\sqrt{2-1/\rho_t}$ and we get
\begin{align*}
    |C_t|/|t|&=|T(q_1,q_2)|/(2-v^2)\\
    &=\frac12\left(1/v-v+1/v-v\right)\left(v+v\right)/\left(2-v^2\right)\\
    &=2-2\rho_t
\end{align*}

Now let $t=\wcTileLD{s}$. We define $\ammaG_t=h_t\cap\set{(x,y)|v\le x\le 1/v}$.
$t$ then consists of exactly one slide $(q_1,q_2)=((v,1/v),(1/v,v))$.
With $|t|=1+\ln((1/v)/v)=1-2\ln(v)\leftrightarrow v=e^{\frac12-\frac{1}{2 \rho_t}}$, we get

\begin{align*}
    |C_t|/|t|&=|H(q_1,q_2)| \rho_t\\
    &=\rho_t \left[\ln(z)+\frac{1}{4}\left(z^2-z^{-2}\right)\right]_{e^{\frac12-\frac{1}{2 \rho_t}}}^{e^{\frac{1}{2 \rho_t}-\frac12}}\\
    &=1-\rho_t(1+\sinh(1-1/\rho_t))
\end{align*}

Note that $t$ is only valid in the sense of generalized tiles. However,
$t$ can be arbitrarily well approximated by a non-generalized tile with area and crown size arbitrarily close to $|t|$ and $|C_t|$.
\end{proof}

%% file: 91-upper-bound-full.tex
\clearpage
\section{Full Proof of the Upper Bound}
\label{sec:full-proof-upper-bound}
In this section, we give the full details for the construction used to show the upper bound in \cref{thm:worst_case_covering}.
Our construction is parametrized by $k$, $A$, and $\varepsilon$.
The main idea is the following (cf.\ \cref{sec:upper-bound}): we construct a point set where \textsc{TilePacking} fails to cover an area larger than roughly $ (1-e^{-2})/2$.

To simplify the analysis, we rotate the unit square $\cU$ clockwise by $45$ degrees and let $\cU_r$ denote the unit square $\cU$ rotated clockwise by $45$ degrees around the origin.
Observe that in $\cU_r$, \textsc{TilePacking} processes the points from right to left.

Recall that our goal is to construct an \emph{$L$-shaped tile} with low density, where each maximal rectangle has roughly the same size $A$.
To this end, we place a large number of points $q_i$ on a hyperbola $h_A$.
We then add for each point $q_i$ on $h_A$ a set of points $p_{i,j}$ with $x(p_{i,j}) = x(q_i) + j\varepsilon$ where $j \in \mathbb{Z}$ and $p_{i,j} \in \cU_r$.
To get a density close to $1/2$, the $y$-coordinates of the points $p_{i,j}$ are chosen as follows:
we place the points $p_{i,j}$ on arcs of functions $f_i$ described by differential equations (see below), where each $f_i$ depends on the two neighboring curves $f_{i-1}$ and $f_{i+1}$.

Observe that \textsc{TilePacking} processes the outer $q_i$ before the inner ones.
This means, once the tile for some outer point $q_i$ has been fixed, the remaining tiles for of all inner points on $f_j$ will be restricted by $t_{q_i}$.

\begin{figure}[t]
\begin{minipage}[t]{0.7\textwidth}
\includegraphics[width=1.0\textwidth]{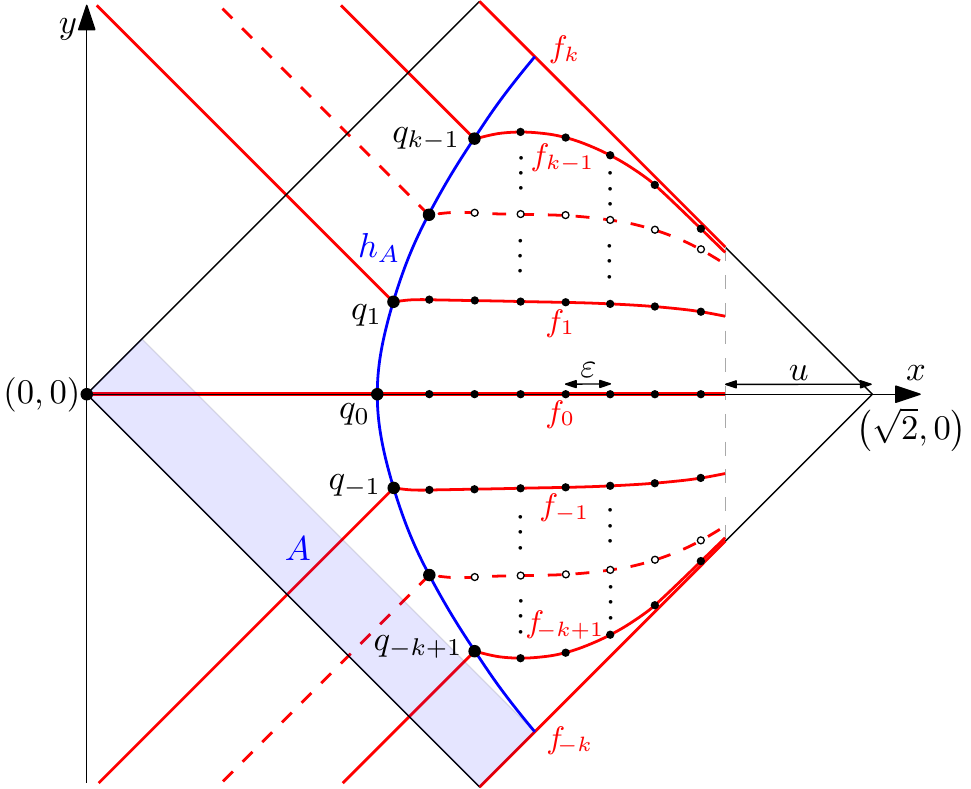}
\caption{Construction with $k = 4$.}
\label{fig:upper_bound}
\end{minipage}
\hfill
\begin{minipage}[t]{0.28\textwidth}
\includegraphics[width=1.0\textwidth]{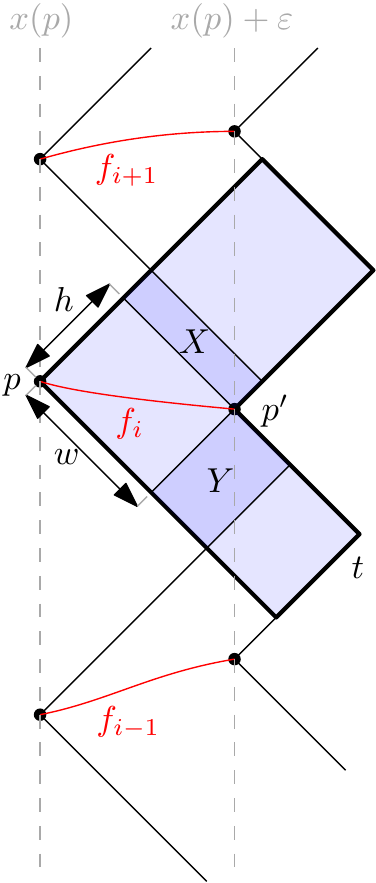}
\caption{Example for a tile created by
$p\in P_{\varepsilon,k}$. For demonstration purposes, the light blue areas are not shown to scale; in truth they are significantly smaller than $X,Y$.}
\label{fig:upper_bound_halfness}
\end{minipage}
\end{figure}

We start by formally defining the point set $P_{\varepsilon,k}$ and the functions $f_i$.
Let $h_A=\set{(x,y) \in \cU_r | x^2-y^2 = 2A }$ be the right branch of a hyperbola centered in the origin that lives in $\cU_r$.
First we define the upper part of the construction with non-negative  $y$ coordintes.
For $i=0,\dots,k-1$, densely choose $k$ points $q_i\in h_A$ such that $y(q_i) \geq 0$ and $\sqrt{2A} = x(q_0) < x(q_1) < \dots < x(q_k)$.
For $i = 0$ and $i = k$, define $f_0(x) = 0$ and $f_k(x)=\sqrt{2}-x$.
For $0 < i < k$, define $f_i: [0,\sqrt{2})\rightarrow \mathbb{R}$ using
\begin{align}
f_i(x(q_i)) &= y(q_i) \label{eq:f-i-intersect-f-a} 
\intertext{ and }
f_i'(x)&=\begin{cases}
-1&\text{ for }x\leq x(q_i)\\
\displaystyle 1 - 2\frac{f_i(x)-f_{i-1}(x)}{f_{i+1}(x)-f_{i-1}(x)}&\text{ for }x>x(q_i). 
\end{cases} \label{eq:f-i-slopes}
\end{align}
This means, each $f_i$ with $0 < i < k$ has slope $-1$ in $\left[0,x(q_i)\right)$, then it intersects $h_A$ in $q_i$ according to \eqref{eq:f-i-intersect-f-a}, and then it has a slope depending on $f_{i-1}$ and $f_{i+1}$ according to \eqref{eq:f-i-slopes}.

For the symmetric part with negative $y$ coordinates we define $q_{-i}=(x(q_i),-y(q_i))$ and $f_{-i}(x)=-f_i(x)$ for $0 < i < k$.
Observe that $q_{-i} \in h_A$ and the $f_{-i}$ adhere to \eqref{eq:f-i-slopes}.
Finally, define $f_{-k} = -f_k$.
We are now ready to define the point set $P_{\varepsilon,k}$ for $\varepsilon>0$ and $k\in\mathbb{N}$ as
\[ P_{\varepsilon,k}= \set{(0,0)} \cup \smash{\bigcup_{i,j \in \mathbb{Z}}} \Set{\big(j\epsilon,\,f_i(j\epsilon)\big) | -k < i < k, x(q_i) \leq j\epsilon < \sqrt{2} } . \]
We assume that $\varepsilon$ and the $q_i$ are chosen such that $\varepsilon$ divides all $x(q_i)$. This implies that $q_i \in P_{\varepsilon,k}$ for all $-k < i < k$.
We now show basic properties for the functions $f_i$.
\begin{lemma}
\label{lem:upper_bound_well_behaved_fi}
Each function $f_i$ intersects $h_A$ exactly once, namely in $q_i$.
Furthermore, $f_i(x)$ is differentiable for all $i=-k,\dots,k$, and $f_{-k}(x)<\dots<f_k(x)$.
\end{lemma}

\begin{proof}
By the choice of the $q_i$, we have $f_{-k}(0)<\dots<f_k(0)$. As long as $f_{-k}(x)<\dots<f_k(x)$, the $f_{i+1}(x)-f_{i-1}(x)$ denominators in the differential equations are positive and therefore the functions are by definition differentiable and thus continuous. Then we can also show that each $f_i$ cannot have more than one shared point with $h_A$: we have
$|f_i'|=|1 - 2(f_i(x)-f_{i-1}(x))/(f_{i+1}(x)-f_{i-1}(x))|\le 1$. W.l.o.g.\ consider the upper branch of $h_A$ with the functional form $h=\sqrt{x^2-2A}$, then we get $|f_i'|\le 1\le x/\sqrt{x^2-2A}=h'$, so there cannot be more than one shared point.

Assume now that the functions are not always ordered, then choose the smallest $\tilde{x}<\sqrt{2}$ such that $f_{i}(\tilde{x})=f_{i-1}(\tilde{x})$ for some $i$. W.l.o.g.\ we can assume
that $f_{i+1}(\tilde{x})>f_{i}(\tilde{x})$ (otherwise choose $i+1$ instead or
consider the symmetric case). Note that such an $i$ must exist,
as otherwise all functions $f_i$ meet, which is impossible since the functions $f_k$ and $f_{-k}$ only meet at $x=\sqrt{2}$.

Define the function $\delta:=f_{i}-f_{i-1}$.
There exists a $v>0$ such that $f_{i+1}(x)-f_{i-1}(x)\ge v$ holds for all $x\le \tilde{x}$, where $f_i$ is well-defined by its differential equation. When evaluating the functions at
$x$, we get
\[
\delta'=f_{i}'-f_{i-1}'=1 -f_{i-1}'- 2\frac{f_i-f_{i-1}}{f_{i+1}-f_{i-1}}\ge-2\frac{\delta}{v}
\]
where the last inequality follows from $f_{i+1}(x)-f_{i-1}(x)\ge v$, as well as the properties of the derivative. We therefore see that $\delta$ decays (at most) exponentially and can never reach $0$, contradicting our assumption of $\delta(\tilde{x})=f_i(\tilde{x})-f_{i-1}(\tilde{x})=0$.
\end{proof}

\begin{lemma}
\label{lem:upper_bound_half_covered_L_tiles}
Let $u,k>0$ and
$\hat{\cU}=(\bigcup_{p\in P_{\varepsilon,k}, x(p)\le \sqrt{2}-u} t_p)\setminus \hat{t}$.
Then \textsc{TilePacking} covers $|\hat{\cU}|/2+c(\varepsilon)$ area
in $\hat{\cU}$ for $P_{\varepsilon,k}$ where
$\lim_{\varepsilon\rightarrow 0}{c_k(\varepsilon)}=0$.
\end{lemma}

\begin{proof}[Proof of \Cref{lem:upper_bound_half_covered_L_tiles}]
By \Cref{lem:upper_bound_well_behaved_fi}, we get
$f_{-k}(x)<\dots<f_{k}(x)$, ensuring that all $f_i$ reside in $\cU_r$.
Now consider a point $p\ne (0,0)$, $x(p)\le\sqrt{2}-u$
that lies on some curve $f_j$ and creates the tile $t$.
Assuming $\varepsilon<u$, there exists another point
$p'=(x(p)+\varepsilon,f_j(x(p)+\varepsilon))\in P_{\varepsilon,k}$.
By \Cref{lem:upper_bound_well_behaved_fi}, we can assume that$f_{i+1}(x)-f_{i}(x)>\varepsilon$ for all $i=-k,\dots,k-1$, $x\le \sqrt{2}-u$. It follows from $|f_i'|\le 1$ that $p'$ is a lower staircase point. (For the same reason there cannot be points further to the right that are lower staircase points.) $t$ is therefore only restricted by the tiles from points with $x$-coordinate $x(p')$, giving it an $L$-shape (see \Cref{fig:upper_bound_halfness}).

\textsc{TilePacking} will choose the larger one of the two maximal rectangles $R_1$ and $R_2$.
Depending on the processing order, \textsc{TilePacking} will process any of the points with the $x$-coordinate $x(p)$ first.
This gives rise to areas that can be either covered by $t$ or the tile of
a point on the neighboring curve. As we will see, these areas are negligible, so w.l.o.g.\ we assume that $t$ covers them.

Since all $f_i$ are differentiable in $[0,\sqrt{2})$, Taylor's Theorem provides the existence of a function $g(x)$ with $\lim_{x\rightarrow 0} g(x)=0$ such that $f_i(x+\varepsilon)=f_i(x)+f_i'(x)\cdot \varepsilon+g(\varepsilon)\cdot \varepsilon$.

Denote by $w,h$ the dimensions of the rectangle $r=R_1\cap R_2$. Then
$w
=(\varepsilon+f_i(x(p)+\varepsilon)-f_i(x(p)))/\sqrt{2}
=(\varepsilon+f'_i(x(p)) \varepsilon+g(\varepsilon)\varepsilon)/\sqrt{2}
=(1+f_i'(x(p))+g(\varepsilon))\varepsilon/\sqrt{2}$ and similarly
$h=(1-f'_i(x(p))-g(\varepsilon))\varepsilon/\sqrt{2}$.

Let $Z$ be the total area of $r$ plus all areas that could have been
covered by neighboring tiles of points with the same $x$-coordinate $x(p)$. It is easy to see that $Z=\LDAUOmicron{\varepsilon^2}$.

$t$ also contains two additional rectangles with an area of
$X=w((f_i(x(p))-f_{i-1}(x(p)))/\sqrt{2}-h)$ and
$Y=h((f_{i+1}(x(p))-f_i(x(p)))/\sqrt{2}-w)$.
Note that this also holds if $x(q_{i\pm 1})>x(p)$, as we extended $f_{i\mp 1}$ with lines of slope $\pm 1$. In this case the two rectangles are restricted by $q_{i\mp 1}$'s tile, respectively.

Hence, when evaluating the functions at $x(p)$:
\begin{align*}
|X-Y|&=|w(f_i-f_{i-1})/{\sqrt{2}}-h(f_{i+1}-f_{i})/{\sqrt{2}}|\\
&=|(1+f'_i+g(\varepsilon))(\varepsilon(f_i-f_{i-1}))/{2}-(1-f'_i-g(\varepsilon))(\varepsilon(f_{i+1}-f_{i}))/{2}|\\
&=|(f_{i+1}(g(\varepsilon)+f'_i-1)-f_{i-1}(g(\varepsilon)+f'_i+1))+2f_i|\cdot \varepsilon/2\\
&=|(g(\varepsilon)(f_{i+1}-f_{i-1})+f'_i(f_{i+1}-f_{i-1})-f_{i-1}-f_{i+1}+2f_i)|\cdot \varepsilon/2\\
&=|g(\varepsilon)| |f_{i+1}-f_{i-1}|\cdot \varepsilon/2\\
&\le|g(\varepsilon)|\cdot \varepsilon \sqrt{2}
\end{align*}
where the last inequality holds by \Cref{lem:upper_bound_well_behaved_fi} and $f_k(x)-f_{-k}(x)\le 2\sqrt{2}$ in $[0,\sqrt{2})$.

W.l.o.g.\ assume $X>Y$.
Then for the tile $t$ with area $|t|=X+Y+Z$, \textsc{TilePacking} covers at most $Z+X\le Z+X/2+(Y+ |g(\varepsilon)|\cdot \varepsilon \sqrt{2})/2=|t|/2+\LDAUOmicron{\varepsilon(|g(\varepsilon)|+\varepsilon)}$.
As $\hat{\cU}$ is the union of such tiles and $|P_{\varepsilon,k}|=\LDAUOmicron{k/\varepsilon}$, we have
a total coverage of $|\hat{\cU}|/2+\LDAUOmicron{k(|g(\varepsilon)|+\varepsilon)}$.
This immediately gives us the function
$c_k(\varepsilon)=\LDAUOmicron{k(|g(\varepsilon)|+\varepsilon)}$
with $\lim_{\varepsilon\rightarrow 0} c_k(\varepsilon)=0$.
\end{proof}

\begin{theorem}
\label{thm:upper_bound}
\textsc{TilePacking} has no better lower bound than
$(1-e^{-2})/2$.
\end{theorem}

\begin{proof}
We analyze the area $\rho$ covered by \textsc{TilePacking} on $P_{\varepsilon,k}$ for some fixed $k$ and $u$ as $\varepsilon$ approaches $0$. The bound then follows from letting
$k$ go to $\infty$ and $u$ go to $0$.

By \Cref{lem:upper_bound_half_covered_L_tiles},
\textsc{TilePacking} covers half of
$\hat{\cU}=(\bigcup_{p\in P_{\varepsilon,k}, x(p)\le \sqrt{2}-u} t_p)\setminus \hat{t}$
(plus $c_k(\varepsilon)$ that approaches $0$ for
$\varepsilon\rightarrow 0$) for each $u>0$.
Additionally, at most $u^2$ area is covered from all tiles at points $p$ with $x(p)>\sqrt{2}-u$.
\textsc{TilePacking} covers at most $A+Q$ area in $\hat{t}$, where at most $Q=\max_i{(x(q_i)-x(q_{i-1}))}+\max_i{(y(q_i)-y(q_{i-1}))}$ is additionally covered due to the $q_i$ points only providing an approximation of $h_A$.
(Note that all $q_i$ lie in $P_{\varepsilon,k}$, so no additional error is introduced.)
In total, using $E=Q+c_k(\varepsilon)+u^2$,
\begin{align*}
\rho&\le A+|\hat{\cU}|/2+E\\
&\le A+(1-|\hat{t}|)/2+E\\
&\le A+(1-(A+\int_{A}^{1}{\frac{A}x\ \mathrm{d} x}))/2+E\\
&\le (1+A+A\ln{A})/2+E
\end{align*}
Minimizing the last term leads to $\rho=(1-e^{-2})/2+E$ at $A=e^{-2}$.
$Q$ approaches $0$ when $k$ goes to $\infty$ since the $q_i$ points lie densely on $h_A\cap \cU_r$. Therefore the error term $E$ approaches $0$ for large $k$ and small $u,\varepsilon$.
\end{proof}